\documentclass[reprint,amsmath,amssymb,aps,prl,floatfix,dvipsnames]{revtex4-2}

\usepackage{graphicx}% Include figure files
\usepackage{dcolumn}% Align table columns on decimal point
\usepackage{bm}% bold math
\usepackage{comment}
\usepackage{amsmath}
\usepackage{amssymb}
\usepackage{graphicx}
\usepackage{xcolor}
\usepackage{mathtools}
\usepackage{blindtext}
\usepackage{hyperref}% add hypertext capabilities
\usepackage[caption=false]{subfig}
\usepackage{txfonts} 
\usepackage{siunitx}
\usepackage{bm}
\hypersetup{colorlinks=true, citecolor=blue, urlcolor=blue, linkcolor=blue}

\newcommand\figref{Fig.~\ref}

\newcommand{\AR}{W}

\begin{document}

\preprint{APS/123-QED}

% \title{Exogenous--endogenous surfactant interaction enables maze solving}
\title{Exogenous--endogenous surfactant interaction yields heterogeneous spreading in complex branching networks} 
% Other titles: Exogenous--endogenous surfactant interaction amplifies heterogeneous spreading in complex branching networks
% Force line breaks with \\

\author{Richard Mcnair$^1$, Fernando Temprano-Coleto$^2$, François J. Peaudecerf$^3$, Fr\'ed\'eric Gibou$^4$, Paolo Luzzatto-Fegiz$^4$, Oliver E. Jensen$^1$, Julien R. Landel$^{1,}$}
{}%Lines break automatically or can be forced with \\
%\author{}%
 \email{Corresponding author: julien.landel@manchester.ac.uk}
\affiliation{%
$^1$Department of Mathematics, University of Manchester, Manchester M13 9PL, UK
\\
$^2$Andlinger Center for Energy and the Environment, Princeton University, Princeton, New Jersey 08544, USA
\\
$^3$Univ Rennes, CNRS, IPR (Institut de Physique de Rennes) – UMR 6251, F - 35000 Rennes, France 
\\
$^4$Department of Mechanical Engineering, University of California, Santa Barbara, California 93106, USA
}

% \author{Fernando Temprano-Coleto}
% %\email{ftempranocoleto@ucsb.edu}
% \affiliation{%
%  Department of Mechanical Engineering, University of California, Santa Barbara
% }%

% \author{Frederic Gibou}
% %\email{fgibou@ucsb.edu}
% \affiliation{%
%  Department of Mechanical Engineering, University of California, Santa Barbara
% }%

% \author{Paolo Luzzatto-Fegiz}
% %\email{pfegiz@ucsb.edu}
% \affiliation{%
%  Department of Mechanical Engineering, University of California, Santa Barbara
% }%

% \author{François Peaudecerf}
% %\email{peaudecerf@ifu.baug.ethz.ch}
% \affiliation{%
%  Institute of Environmental Engineering, Department of Civil, Environmental and Geomatic Engineering, ETH Zurich 
% }%

% \author{Oliver E. Jensen}
% %\email{oliver.jensen@manchester.ac.uk}
% \affiliation{%
% Department of Mathematics, University of Manchester.
% }%

% \author{Julien R. Landel}
% %\email{julien.landel@manchester.ac.uk}
% \affiliation{%
% Department of Mathematics, University of Manchester.
% }%


\date{\today}% It is always \today, today,

% \begin{abstract}
% % Character count: 626/600 (over the limit, but may still be fine)
% Experiments showed that surfactant added at the inlet of a liquid maze can find the solution path. We reveal how the maze-solving dynamics arise from interactions between the added surfactant and natural surfactant present at the liquid surface. We simulate the dynamics using a nonlinear diffusion model solved with a discrete mimetic scheme on a branching network. Natural surfactant transforms a local spreading problem into a non-local problem with an omniscient view of the maze geometry, key to the maze-solving dynamics. Our results offer insight into surfactant-driven transport in complex networks such as lung airways.
% \end{abstract}

\begin{abstract}
% Character count: 596/600 (now under the limit, prev version above)
Experiments have shown that surfactant introduced to a liquid-filled maze can find the solution path. We reveal how the maze-solving dynamics arise from interactions between the added surfactant and endogenous surfactant present at the liquid surface. We simulate the dynamics using a nonlinear model solved with a discrete mimetic scheme on a graph. Endogenous surfactant transforms local spreading into a non-local problem with an omniscient view of the maze geometry, key to the maze-solving dynamics. Our results offer insight into surfactant-driven transport in complex networks such as lung airways.
\end{abstract}

\maketitle

%\tableofcontents

\refstepcounter{equation}\label{eq:start}
% \jl{[INTRODUCTION (1 to 1.5 page)]}

% \justifying

\begin{figure}[t]
    \centering
    \captionsetup{farskip=0pt, nearskip=0pt,skip=0pt}
    \includegraphics[width=\linewidth]{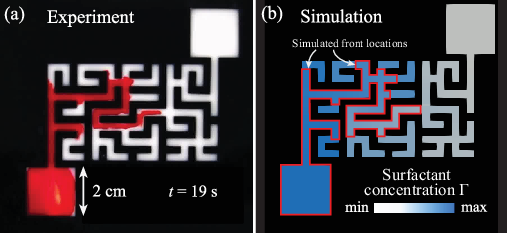}
    % \includegraphics[width=\linewidth]{Figure1v6shaded.eps}
    % \label{fig:Figure1a}
    \caption{\label{fig:experiment} 
    (a) Photograph of the liquid maze experiment \cite{temprano2018soap}:  surfactant deposited at the inlet finds the solution path with minimal penetration into lateral branches, as shown by red dye advected by Marangoni forces. (b) Results from our  model \eqref{eq:NonlinearDiff}  showing the concentration (blue-scale) of the  surfactant field in the same geometry; the red front is the exogenous-endogenous boundary.  %The model simulation was computed for an initial endogenous--exogenous surfactant concentration ratio $\delta=0.15$. 
    %Comparison of the maze experiments \cite{temprano2018soap} with results from our model. (a) Photograph of the experiment showing the motion of a red dye front in a maze  filled with liquid (milk) due to the introduction of exogenous surfactant at the inlet. The surfactant carries the dye front through the solution path to the outlet within approximately 1 minute, with minimal penetration in the side branches. In (b) we capture this behaviour in our model simulation of the experiment at $t=5340$ with $\delta=0.15$ ($t$ and $\delta$ are defined in the text).
    \vspace{-7pt}
    }
\end{figure}

% \jl{[Context and general problem]}

% Surfactant can modify the dynamic boundary condition relating the bulk stresses in the liquid and gas phases through the so-called Marangoni effect which produces interfacial stresses from surfactant concentration gradients. When surfactant are endogenous to a system,  since they are not visible through the naked eye. They can yield often unexpected but measurable macroscopic changes on the flow dynamics. Canonical examples of the unexpected impact of endogenous surfactant have been known for a long time, with the impact on the drag of rising air bubbles in water, whereby surfactants could change the nominally free stress interfacial condition, without surfactant, to a no-slip or intermediate complex stress balance, with surfactant. 

Recent experiments \citep{temprano2018soap} showed that surfactant, made visible by dye, can  `solve' a liquid maze (\figref{fig:experiment}(a), Supplementary Movie 1~\cite{Svideo}). The surfactant propagates along the solution path, with  little advance into lateral branches. This phenomenon seems counterintuitive, as no  mechanism appears to draw the surfactant preferentially to the maze exit: one might expect Marangoni stresses to spread the  surfactant throughout the whole maze. Yet, the surfactant proceeds with apparent awareness of the whole geometry. 
Although no quantitative theory exists, a hypothesis was suggested based on the potential effect of pre-existing, `endogenous’ surfactant in the maze liquid  \citep{temprano2018soap}. The spread of the  `exogenous’ (added)  surfactant would be influenced by the endogenous surfactant, in a manner sensitive to the maze geometry. 
Inspired by these observations, we investigate the impact of \emph{endogenous} surfactant on the spreading of \emph{exogenous} surfactant on thin liquid layers confined in  asymmetric branching networks.

% Inspired by recent experiments from Temprano-Coleto et al. \citep{temprano2018soap}, we investigate the impact of \emph{endogenous} surfactant on the spreading of \emph{exogenous} surfactant at the surface of thin liquid layers confined in asymmetric branching networks. By depositing exogenous surfactant at the surface of a liquid filling a maze, the surfactant was observed to `solve' the maze with minimal penetration into side branches, as revealed by the passive red dye in \figref{fig:experiment}(a) (see Supplementary Movie~\cite{Svideo}). This may seem counterintuitive, because one might expect surfactant-induced Marangoni stresses to spread the exogenous surfactant throughout the whole maze and not preferentially choose the solution path, as no visible mechanism draws the flow to the maze exit, yet the dyed surfactant proceeds as if it had awareness of the overall geometry. Temprano-Coleto et al.~\citep{temprano2018soap} hypothesized that endogenous surfactant, originally present at the liquid surface %and the asymmetry 
% of the maze, could potentially explain the maze-solving behaviour. However, this hypothesis remains untested, as there are presently no quantitative models of the maze-solving dynamics.

Whether from natural or contaminant sources, endogenous surfactants are usually undetectable \textit{a priori}, but have   macroscopic influence on flow dynamics across numerous applications \citep{peaudecerf17,Hourlier-Fargette17,manikantan2020surfactant,koleski20,faugaret20,temprano23}. %Canonical examples include the increase in drag of air bubbles rising in water contaminated with surfactant (citations). Recently, the effect of endogenous surfactant has been observed across a number of  flow phenomena, suggesting a widespread impact. Endogenous surfactant can deteriorate the drag reduction performance of superhydrophobic surfaces (citations), reduce the speed of sliding drops and bubbles along soft substrates used in microfluidic and biomedical applications (cite hourlier fargette), affect the stability and morphology of convective surface flows induced by rotation \citep{faugaret20,faugaret22} or thermocapillary forcing \cite{koleski20}. %and prompting Manikantan and Squires to refer to them as ``hidden variables controlling fluid flows'' \cite{manikantan2020surfactant}. 
% Trace amounts of endogenous surfactants can severely reduce the drag reduction performance of superhydrophobic surfaces in laminar flows \citep{peaudecerf17}. They can reduce the speed of sliding drops and bubbles along soft substrates, such as silicone elastomers or gels found in microfluid and biomedical applications (cite Hourlier Fargette). They have also been shown to affect the stability and morphology of convective surface flows induced by rotation \citep{faugaret20,faugaret22} or thermocapillary forcing \cite{koleski20} \jl{[any more examples/citations?]}.
% In all these examples, Marangoni forces arise due to surfactant concentration gradients generated by steady external forcing in confined geometries. In the maze experiments, however, the flow is itself generated by unsteady Marangoni stresses through the addition of a finite amount of exogenous surfactant. Therefore, understanding the impact of the endogenous surfactant on the exogenous surfactant dynamics requires a time-dependent analysis, in which the complex asymmetric tree-like geometry of the maze plays a key role.
Natural endogenous surfactant in human lungs  can affect exogenous surfactant  therapies \citep{espinosa93,grotberg1995interaction,halpern98,zhang03,stetten2018surfactant,garcia-mouton23} for various diseases, including acute respiratory distress syndrome (ARDS) \citep{anzueto96}. %, bronchopulmonary dysplasia \citep{cpapvssurfactant10} and bacterial pneumonia  \citep{baer2019new}. 
Although  efficacy of surfactant replacement therapy is indisputable for neonatal respiratory distress syndrome \cite{avery59,jobe1993pulmonary,rodriguez2003management,halliday2008surfactants}, there remains debate for other pathologies \citep{baudouin04,zuo08,sankar16,veldhuizen21}, including ARDS, which has a 40\% mortality rate in the US \citep{rubenfeld05}. To understand mixed results from clinical trials \citep{baudouin04}, experimental and theoretical models analysed  physical, chemical and biological  processes in  surfactant transport through branching networks simulating lungs \cite{levy14,stetten2018surfactant,baer2019new}. In distal lung airways, where surfactant therapy is  effective, Marangoni transport dominates gravitational transport \cite{halpern98,espinosa99,halpern98,grotberg2001respiratory,stetten2018surfactant}.
Endogenous surfactants
% , produced by the lung, previously administered or from contaminant sources, 
can affect therapeutic surfactant transport, preventing it from reaching distal airways \citep{jensen1992insoluble,espinosa93,grotberg1995interaction,halpern98,bull99,zhang03,garcia-mouton23}. 

Pulmonary surfactant models have used simplifying geometric assumptions to  reduce analytical, computational and experimental difficulties. Single-airway models \citep{baer18,garcia-mouton23} may suppress crucial dynamics associated with branching \citep{zhang03,kazemi19}. More complex models used a Weibel morphometry \citep{weibel62}, which yields a dichotomous, symmetric and self-similar branching network \citep{espinosa93,grotberg1995interaction,halpern98,espinosa99,zhang03,filoche15,copploe19}; 
%In many  models \citep{espinosa93,grotberg1995interaction,halpern98,espinosa99,zhang03,filoche15,copploe19}, the lung  geometry was modelled using a Weibel morphometry \citep{weibel62}, which yields a dichotomous, symmetric and self-similar branching network. Other studies have also simulated transport into lungs using only a single branch model \citep{baer18,garcia-mouton23}. Such simplifying geometric assumptions vastly reduce analytical, computational and experimental difficulties. 
however, human lungs are  asymmetric and highly irregular \citep{horsfield68,tawhai04,schmidt04}. Recent models \citep{kazemi19,elias-kirma21}, which neglected Marangoni spreading, showed that transport by air flow and gravity in asymmetric lungs can yield heterogeneity in surfactant distribution.
%However, they did not study surfactant Marangoni spreading.
% \jl{[Now is probably the most tricky part to get right, as it will probably be the most contentious for lung experts who may argue we're a bit fraudulent here in the connection... I propose to be as honest and transparent as possible, but obviously this could also backfire from the eyes of PRL editors/reviewers as being too weak!]} 
% The primary purpose of the maze experiments conducted by Temprano-Coleto et al was not to model surfactant transport in lungs. 
% Nevertheless, the strong asymmetry shown by the exogenous surfactant as it solves an asymmetric maze may provide important insight about possible sources of heterogeneities in the surfactant transport through  realistic lung models.
The maze experiments of \cite{temprano2018soap} were not designed to model surfactant transport in lungs; nevertheless, the two problems share key underlying physical features. %enabling   surfactant to solve  mazes is similar to  transport of  surfactant in distal lung airways. %Marangoni forces drive surfactant through an asymmetric branching network against viscous forces from a thin liquid film. 
Understanding surfactant dynamics in  mazes can give insight about heterogeneities affecting surfactant spreading in large   complex networks, such as human lungs.

To test the hypothesis that endogenous surfactant explains the maze-solving behaviour of exogenous surfactant \citep{temprano2018soap}, and provide new physical insight, we model surfactant transport on a thin film confined in a 1D branching network, replicating the maze geometry from \citep{temprano2018soap}. 
The experimental maze consists of a  network of grooves in an acrylic substrate, with larger 20 x 20 mm  branches at the inlet and outlet (\figref{fig:experiment}(a)). The large outlet creates an  asymmetry in the geometry. Branches were filled with a 50\% milk-cream mixture, which provides a  dense, viscous liquid contrasting with red tracer dye added at the inlet. A surfactant drop  (0.3\% w/w soap--water solution) was deposited at the inlet  at $t=0$. Dye is buoyant in the milk, and therefore remains at the surface. Within  60\,s, one of the red dye fronts, assumed to follow the exogenous surfactant fronts, reached the outlet, through the solution path of length 155 mm (Sec.~S1 \citep{SM}).
We assume the dynamics are governed by a balance of viscous and Marangoni stresses, that solubility and diffusion of surfactant are negligible, and that inertia and interface deformation are negligible, as shown quantitatively in \citep{SM}. Endogenous and exogenous contributions are modeled as a single surfactant concentration field $\Gamma$.
The axial and vertical coordinates are $x$ and $z$, $t$ is time, $u|_s$ the surface axial velocity, and $\gamma$ the surface tension. 
Variables are normalized by appropriate physical scales \citep{SM}, such that e.g.~$0\leq z \leq 1$.
The key equations expressing surfactant conservation and stress balance at the surface are therefore, in nondimensional variables, 
\begin{equation}
\Gamma_t + (u|_s \Gamma)_x = 0, \qquad \text{and} \qquad  \gamma_x = u_z|_s,
\tag{\theequation a,b}\label{eq:start_ab}
\end{equation}
where subscripts in $x,z,t$ denote partial derivatives, and $u_z|_s$ is the viscous stress at the surface.  To introduce key ideas, ignore initially the effect of lateral ($y$) confinement, such that to leading order the viscous flow in the interior is governed by $u_{zz} = p_x$. Integrating twice in $z$ and assuming that the volume flux along the channel is uniformly zero, such that $\int_0^1 u \,dz = 0$, one finds the surface velocity $u|_s = p_x/6$ and viscous stress $u_z|_s = 2 p_x/3$. Eliminating $p_x$ gives $u_z|_s = 4 u|_s$. The nondimensional surfactant concentration and surface tension are related by $\gamma_x = -\Gamma_x$ \citep{SM}; substituting this relation, together with $u_z|_s = 4 u|_s$, into (\ref{eq:start}b) yields $u|_s = -\frac{1}{4}\Gamma_x$. Eliminating $u|_s$ in (\ref{eq:start}a) yields an equation that only involves $\Gamma$,
%
%, and substituting into  we find $\gamma_x = 4u|_s$. ; substituting in (\ref{eq:start}b), and eliminating $u|_s = \frac{1}{4}\Gamma_x$ from (\ref{eq:start}a) yields
%
\begin{equation}
\Gamma_t = \tfrac{1}{4}(\Gamma \,\Gamma_x)_x \equiv \tfrac{1}{8}(\Gamma^2)_{xx}.
\label{eq:NonlinearDiff}
\end{equation}
An alternative derivation of (\ref{eq:NonlinearDiff}), starting from lubrication theory, is possible \citep{SM}.  If we include the effect of lateral no-slip boundaries at $y = \pm \AR/2$, the factor of 1/4 in (\ref{eq:NonlinearDiff}) and $u|_s$ is replaced by a positive coefficient $f(\AR)<1/4$, calculated \textit{a priori} as a function of $\AR$. This factor is a nonessential time scaling, as we rescale $t$ by experimental completion time; however, this analysis shows that narrow channels significantly reduce the surface velocity, helping to keep the flow within the viscous limit (for \citep{temprano2018soap} $\AR\approx 0.94$ yielding $f\approx 0.15$).

%as explained further below.

\begin{figure}
	\centering
 \captionsetup{nearskip=0pt, farskip=0pt, skip=0pt}
	% ~~~~~~~\includegraphics[width=0.485\textwidth,trim={9mm 2mm 0mm 28mm},clip]{ModelFigurePossibility17.eps}
 \includegraphics[width=\linewidth]{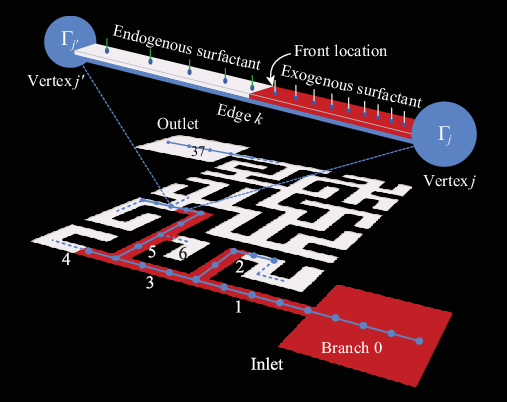} %trim={left bottom right top},trim={4pt 6pt 9pt 4pt}
	\caption{
	Schematic showing how the  transport model \eqref{eq:NonlinearDiff} is implemented in a maze mimicking the experiment, with the branch numbering scheme indicated (Fig. S1 \cite{SM}). The maze topology is modelled by a 1D branching network (blue). Along each edge, the model  results are used to advect the exogenous surfactant fronts (red). %At each vertex, continuity of surfactant concentration and flux are imposed. No-flux is imposed at dead ends. To represent the initial conditions of the experiments, the inlet branch constitutes a reservoir with a finite amount of exogenous surfactant (red) set at concentration $\Gamma=1$ at $t=0$, whilst the rest of the maze is filled with endogenous surfactant (white) with a uniform concentration set to $\Gamma=\delta$ at $t=0$, where $\delta$ is the main fitting parameter. The red dye fronts at the exogenous--endogenous surfactant boundaries are simultaneously tracked in a Lagrangian way through the maze.
 \vspace{-12pt}
 }
	\label{fig:Schematic}
\end{figure}

Model \eqref{eq:NonlinearDiff} is solved numerically in a branching network simulating the experimental maze topology, illustrated in \figref{fig:Schematic}. 
%The network consists of 1D branches following the maze topology. 
At  junctions, we impose continuity of surfactant concentration and  flux, whilst no flux is imposed at  dead-ends (Sec.~S2 \citep{SM}). The dye fronts are computed by Lagrangian integration of the surface velocity $u|_s$. %$u|_s=-\Gamma_x$ produced by the surfactant field, obtained from \eqref{eq:NonlinearDiff}, understood as surface transport of the form $\Gamma_t +(\Gamma u_s)_x= 0 $. %, at the location of each front.
To model the initial conditions, we assume that exogenous surfactant is deposited instantaneously and uniformly at the inlet at $t=0$. The normalised initial concentration is set to $\Gamma=1$ at the inlet, whilst the rest of the maze has initial endogenous surfactant concentration $\Gamma = \delta$ everywhere, where $0 < \delta < 1$ is an empirical parameter. % the main empirical fitting parameter in our model. 
% The initial exogenous--endogenous concentration difference $1-\delta$ is the source of the Marangoni stresses which drive the time-dependent dynamics in the maze experiments, until the surfactant concentration field reaches equilibrium as $t\to\infty$.
%
The branches are numbered as illustrated in Fig.~\ref{fig:Schematic} (see \citep{SM} for further details). Branch $i$ of length $L_i$ (normalised by the solution path  length) %$0\leq i\leq 37$ 
has longitudinal coordinate $0\leq x_i \leq L_i$. Internal branches have lengths representative of the experiment. The greater width of the inlet ($i=0$) and outlet ($i=37$) branches is treated by constructing 1D equivalents. Since the outlet has small variations in concentration, its length $L_{37}$ is calculated by dividing its surface area by the width of an interior branch. The inlet length $L_0$ is calculated by matching the amount of initial exogenous surfactant.

%Transport in the outlet  is assumed to be 1D, owing to  small variations in concentration in this branch. Its length $L_{37}$ is calculated by dividing the surface area of the outlet by the width of a narrow branch.
%Transport at the inlet is  complex due to large variations in concentration during the experiment. We model it as a 1D branch, with its length representing the total amount of exogenous surfactant released initially. 
We estimate the ratio of endogenous to exogenous surfactant mass from the {equilibrium} state of the experiment, as encoded by the ratio of white versus red surface areas (Sec.~S2.4 \citep{SM}). To avoid numerical instabilities, care is taken to regularize the initial surfactant concentration distribution between the inlet and the first branch (Sec.~S2.2 \citep{SM}).
The time at which the front enters  branch $1$ is labelled as $t=0$. As noted earlier, in the model, we re-scale time so that the completion time (when the front enters the outlet) matches the experiment up to a multiplying factor $\tau_{end}$ chosen between $0.5 \leq \tau_{end}\leq 1.5$.

To solve the  system of equations of the form \eqref{eq:NonlinearDiff}, coupled between each of the branches of the maze, we use a mimetic finite-difference numerical method, which balances fluxes at junctions to machine precision. Mimetic finite-difference methods \cite{tikhonov1962homogeneous,van2017modeling, hyman1997natural} use tools from graph theory %, differential forms 
and discrete calculus to obtain finite-difference operators which reproduce desired features of vector % and tensor 
calculus operators.  As shown in Fig.~\ref{fig:Schematic}, we represent the maze as a graph with vertices at every junction and dead end. Extra vertices are  placed along each edge to refine the discretisation. We construct finite-difference mimetic operators on this graph using the incidence matrix of the network $\mathsf{A}$, the columns of which correspond to the vertices of the graph, and the rows of which correspond to the edges. Each edge is assigned an orientation directed away from the maze inlet. Components of $\mathsf{A}$ are zero except at vertices where an edge leaves ($-1$), and  where an edge enters ($+1$). Metric information is encapsulated by the diagonal edge and vertex length matrices $\mathsf{L}$ and $\mathsf{V}$ respectively. Each component of $\mathsf{L}$ is the length of an edge, and each component of $\mathsf{V}$ is one half the sum of lengths of every edge incident to the corresponding vertex --- an important nuance ensuring mass conservation (see Sec.~S3.1 \citep{SM}). The matrix $\mathsf{L}^{-1}\mathsf{A}$ is therefore a finite difference gradient operator acting on scalar functions defined on the graph's vertices, and is second order accurate at edge midpoints. Likewise, $-\mathsf{V}^{-1}\mathsf{A}^T$ is a divergence operator acting on functions defined at edge midpoints, and is second order accurate at the vertices. The concentration of surfactant at the vertices of the graph form the components of vector $\boldsymbol{\Gamma}(t)$. The system of differential equations of the form \eqref{eq:NonlinearDiff} solved on the graph is found by combining gradient and divergence operators into a Laplacian, and is therefore
% $
%     \mathrm{d} \boldsymbol{\Gamma}/\mathrm{d}t  = f(\AR)\mathsf{V}^{-1} \mathsf{A}^T \mathsf{L}^{-1} \mathsf{A} \boldsymbol{\Gamma}^2/2,
% $
\begin{equation}
\frac{\mathrm{d} \boldsymbol{\Gamma}}{\mathrm{d} t}  =  -\frac{f(\AR)}{2} \mathsf{V}^{-1} \mathsf{A}^T \mathsf{L}^{-1} \mathsf{A}\boldsymbol{\Gamma}^2,
\end{equation} 
with $\boldsymbol{\Gamma}^2$  the component-wise square of  $\boldsymbol{\Gamma}$. 
This mimetic finite-difference formulation automatically implements no-flux boundary conditions for all bounding vertices of the graph (dead ends), and continuity of concentration and flux for all junctions within the graph. A semi-implicit scheme solves this system of equations in time (Sec.~S3.2 \citep{SM}).
Simultaneously, we solve  the equation tracking the  dye fronts, which in branch $i$ at $x_i = \Lambda_i(t)$ is
$
\partial \Lambda_i/\partial t = u|_s(\Lambda_i,t)  = - f(\AR) \,\Gamma_x(\Lambda_i,t)
$.
%, which can be found from consideration of the analysis presented in \cite{grotberg1995interaction}. 
When a front reaches a junction,  new fronts are  created in  the downstream  branches.

\begin{comment}
To solve the  system of equations of the form \eqref{eq:NonlinearDiff}, coupled between all  the branches, we use a mimetic finite-difference numerical method, which balances fluxes at junctions to machine precision. Mimetic finite-difference methods \cite{tikhonov1962homogeneous,van2017modeling, hyman1997natural} use tools from graph theory %, differential forms 
and discrete calculus to obtain finite-difference operators which reproduce desired features of vector % and tensor 
calculus operators.  We represent the maze as a graph with vertices at  junctions and dead-ends (Fig. \ref{fig:Schematic}). Additional vertices are placed along each edge for refinement. We construct a mimetic finite-difference approximation of \eqref{eq:NonlinearDiff} on the graph and use
% $
%     \mathrm{d} \boldsymbol{\Gamma}/\mathrm{d}t  = f(\AR)\mathsf{V}^{-1} \mathsf{A}^T \mathsf{L}^{-1} \mathsf{A} \boldsymbol{\Gamma}^2/2,
% $
a semi-implicit scheme to solve the resulting system in time (Sec.~S3.2 \citep{SM}).
Simultaneously, we solve  the equation tracking the red dye fronts which, for a front in branch $i$ at $x_i = \Lambda_i(t)$, is
$
\mathrm{d} \Lambda_i/\mathrm{d}t  = -  \Gamma_x(\Lambda_i,t)
$.
%, which can be found from consideration of the analysis presented in \cite{grotberg1995interaction}. 
When a front reaches a junction,  new fronts are  created in  the downstream  branches.
\end{comment}

% The linear diffusion model can significantly reduce computational costs of simulating transport in large and complex networks under small concentration gradients.
To  reduce computational costs and provide additional insight, we consider  the limit of small concentration gradients, which arises in applications where  concentration differences between exogenous and endogenous surfactants are small. We  decompose the concentration into a time- and space-averaged component and a fluctuating component: $\Gamma = \bar{\Gamma}+ \hat{\Gamma}(x,t)$, with $\vert \hat{\Gamma}\vert \ll \bar{\Gamma}$. Model \eqref{eq:NonlinearDiff} simplifies to the linear diffusion equation $\hat{\Gamma}_t = f(\AR)\bar{\Gamma} \hat{\Gamma}_{xx}$. In the discrete mimetic representation, % we can show that 
the Laplacian operator $f(\AR)\bar{\Gamma}\mathsf{V}^{-1} \mathsf{A}^T \mathsf{L}^{-1} \mathsf{A}$ has orthogonal eigenvectors $\boldsymbol{\phi}_n$ under an inner product $\langle \phi_l, \cdot \phi_m \rangle = \phi_l^T \mathsf{V}\phi_m$ (Sec.~S4 \citep{SM}). The vector of vertex concentrations can be decomposed in a sum of eigenmodes, namely $\boldsymbol{\Gamma}(t) = \sum_{n=1}^{N} A_n\boldsymbol{\phi}_n e^{-\lambda_n t}$, with amplitudes $A_n= \langle \boldsymbol{\phi}_n, \boldsymbol{\Gamma}(t=0)\rangle$; $\lambda_n$ are the corresponding eigenvalues. In the linear regime, the concentration in the maze can be approximated without the need for time integration, by instead using only a finite number $N$ of eigenmodes.

% \jl{[RESULTS  (1.5 pages)]}
% \jl{[How are the model results compared with the experimental results]}
\begin{figure}
		\centering
  \captionsetup{skip=0pt}
		\includegraphics[width=\linewidth]{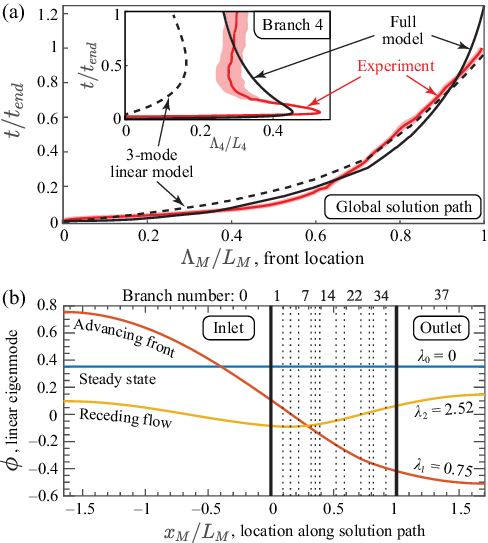}

		\caption{(a) Front location $\Lambda_M(t)$ in time along the solution path, normalised by overall length $L_M$, from experiments (red curves, experimental error shaded pink), nonlinear model (black curves), and 3-mode linear model (dashed curves). Inset shows lateral branch $4$. %, normalised with the branch 4 length $\Lambda_4/L_4$ 
        %(see Fig.~S5 for other branches \citep{SM}). %Model predictions have been fitted, such that $\delta\approx 0.15$ minimizes the error between the model and experiment. 
        % Black dashed lines show the linear model with the first $3$ eigenmodes. Vertical axes are normalised by $t_{end}$, the experimental completion time.
		(b) First 3 eigenmodes of the linear model along the solution path, with eigenvalues $\lambda$ shown. %, with normalised coordinate $x_M/L_M$.  
         %The three modes reproduce qualitatively the maze-solving dynamics and the receding behaviour in lateral branches. 
        Thick vertical black lines show the start and end points. %, with the components to the left and right representing the inlet and outlet branches $0$ and $37$, respectively. 
         Dashed lines show junctions.  %Eigenvector fields along lateral branches are not shown. 
        \vspace{-10pt}
        }
        \label{fig:Results}
\end{figure}

We test our models by comparing their predictions to the dye front locations in the experiments, as shown in Fig.~\ref{fig:Results}(a), which plots time versus location along the global solution path; the inset shows data for a lateral branch. %The nonlinear model results are plotted by black continuous lines. 
The nonlinear model (shown by the black continuous line) captures the spreading dynamics. Its least-square error relative to the experiment is minimized by setting the order-one coefficients $\delta = 0.15$ and $\tau_\text{end}=0.24$ \citep{SM}. %for the nonlinear model, calculated across the global solution path as well as the lateral branches, is minimised by $\delta = 0.15$ and $\tau_\text{end}=0.24$ \citep{SM}. 
In the experiments, fronts in lateral branches are observed to eventually recede after initially advancing; this is most visible near the inlet (Supplementary Movie~\cite{Svideo}); an example is shown in the inset of Fig.~\ref{fig:Results}(a). This receding effect occurs as the concentration at the beginning of the lateral branch eventually decreases over time, as the exogenous surfactant spreads across the rest of the maze, yielding a reversal of the Marangoni stress. The nonlinear model captures the flow reversal qualitatively, generally overpredicting relaxation time (inset of Fig.~\ref{fig:Results}(a)).
The nonlinear model predictions are not  sensitive to the exact value of $\delta$: any value in the range $0.02\leq \delta \leq 0.21$ captures the  dynamics in the experiment. %The value  $\delta\approx 0.15$ yields quantitative agreement along  the solution path and qualitative agreement in the lateral branches, and is consistent with expectations from  experimental conditions.
% The compression of endogenous surfactant occurs in lateral branches ahead of the \textit{local} exogenous surfactant fronts,  after the front through the main solution path has passed, as revealed by our model (\figref{fig:experiment}(b), fronts in red and concentration field in blue scale, see also supplementary videos~\cite{Svideo}). Indeed, the concentration gradient, shown by the colour gradient, in the lateral branches nearer to the inlet has vanished.
 %In contrast,  the concentration gradient throughout the solution path is still visible, at this intermediate time in the experiment.
% We plot the exogenous--endogenous concentration field in the maze in \figref{fig:experiment}(b), at a time corresponding to the experimental picture in \figref{fig:experiment}(a) and with $\delta=0.15$.  

%The linear  model reduces the computational cost of simulating transport in large complex networks under small concentration gradients.
Since the shapes of the eigenmodes depend only on the network geometry, transport dynamics in branching networks can be studied from the dominant eigenmodes and eigenvalues, without having to solve the full nonlinear transport model. The maze-solving dynamics described above are already reproduced by a linear model truncated to just three modes, as shown by the dashed lines in \figref{fig:Results}(a). The front location on the global solution path is captured quantitatively. The spatial structure of each mode is in \figref{fig:Results}(b).
The zeroth mode represents the surfactant concentration at steady state, $t\to\infty$. The first mode, $\lambda_1=0.75$, has a  decreasing profile along the solution path, corresponding to the main concentration gradient driving the front, key to the maze-solving behaviour. The second mode, $\lambda_2=2.52$, presents a wavy profile, which, in combination with the first mode, controls the receding dynamics observed in  lateral branches. %Higher-order modes (not shown here) capture more finely the dynamics along the solution path and in the lateral branches. %However, only the first 3 modes are needed to capture qualitatively the 1D dynamics in the maze. 
The exact temporal evolution requires computation of the mode amplitudes, which depend on initial conditions. 
If three modes were used, $\delta\approx0.03$ minimized the least-square error. 
%A value $\delta\approx0.03$ was found to optimize the linear model  against the experiment, if only 3 modes were used. 
Using a larger number of modes, e.g. of order 100,  recovers  $\delta\approx0.15$ found by the nonlinear model. The main challenge for the linear model is to capture the nonlinear distribution of concentration at $t=0$, better approximated using a larger number of modes.

\begin{figure}[t]
    \centering
    \captionsetup{farskip=0pt, nearskip=0pt, skip=0pt}
    \includegraphics[width=\linewidth]{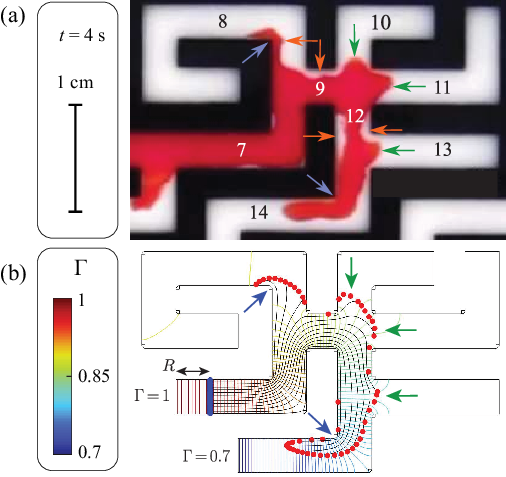}
     % \includegraphics[width=0.95\linewidth]{ExpImageForPaper5.eps}
     % % \includegraphics[width=0.95\linewidth]{ExpImageForPaper6.eps}
     % \includegraphics[width=0.95\linewidth]{ExpImageForPaper4.eps}
    %  \label{fig:ExpImageForPaper1}
    % \includegraphics[width=\linewidth]{MiddleBranchFig7.eps}
    % \label{fig:MiddleBranchFig3}
    \caption{\label{fig:2DDynamics} (a) Experiment, showing branches 7--14. (b) 2D nonlinear simulation.
    % Boundary conditions are no-flux of surfactant at every boundary apart from the inlet and outlet with Dirichlet conditions as shown. A small initial concentration profile of half a cosine wave over the length $R$ connects $\Gamma=1$ to $\Gamma= 0.7$ which is the uniform initial condition everywhere else. 
    Tracers (red dots) are advected from their starting position (blue dots) along Lagrangian  trajectories (black lines). Experimental features are reproduced qualitatively: the front spreads inside sharp corners (blue arrows) and centrally at straight  junctions (green arrows), owing to the bending of the concentration contours. 
    \vspace{-8pt}}
\end{figure}

To assess the importance of 2D dynamics neglected in the model, we perform a 
%test the ability of our model \eqref{eq:NonlinearDiff} to simulate the experiments, we compare 2D features of the spreading fronts with a 
2D simulation of the surfactant transport through branches 7--14 (see \figref{fig:2DDynamics}). We use COMSOL\textsuperscript{\textregistered} to solve the 2D analogue of the 1D model \eqref{eq:NonlinearDiff}, namely $\Gamma_t = \boldsymbol{\nabla}  \cdot (\Gamma \boldsymbol{\nabla}\Gamma) /4$, in the maze geometry. 
A good match to experiment is achieved by imposing $\Gamma =1$ at the entrance of branch~7 and $\Gamma=0.7$ at the exit of branch~14, with initial condition $\Gamma=0.7$ everywhere; as before, care is taken to regularize the initial $\Gamma$ field \citep{SM}.
%
%A concentration difference of $0.3$ across this maze section gives the best fit between simulation and experiment. Hence, we impose boundary conditions $\Gamma=1$ and $\Gamma=0.7$ at the entrance and exit of branches 7 and 14, respectively, continuity at junctions and no-flux at dead-ends. 
%(Sec.~S8 \citep{SM})
%Initial conditions are $\Gamma=0.7$ everywhere, except a half cosine profile connecting $\Gamma=1$ at the branch 7 inlet to $\Gamma=0.7$ over a short length $R$ (\figref{fig:2DDynamics}(b)). 
%
We use the particle-tracing module to track the exogenous--endogenous surfactant front. Dominant dynamical features are reproduced by the 2D model (\figref{fig:2DDynamics}). In the experiments, the exogenous surfactant tends to spread  along the inside  of sharp corners (blue arrows), whilst it spreads through the middle of the path as it enters new branches from a junction (green arrows). The 2D simulation captures qualitatively these behaviours, which are  due to the compression of  surfactant concentration level curves at inside corners (\figref{fig:2DDynamics}(b)). The nonlinear  model appears consistent with surfactant dynamics on 2D confined surfaces. However, the 2D simulation does not capture some of the gaps that appear between the maze walls and dye (orange arrows). These features could be due to aspects omitted from the model such as geometrical imperfections, surface deformation at menisci or chemical heterogeneities.% or other sources of experimental noise.
	
% 	\begin{figure}[]
% 		\begin{subfigure}{0.46\textwidth}
% 		\includegraphics[width=\textwidth]{ExpImageForPaper1.png}
% 		\caption{}
% 	\end{subfigure}
% 		\begin{subfigure}{0.48\textwidth}
% 		\includegraphics[width=\textwidth]{MiddleBranchFig3.eps}
% 		\caption{}
% 	\end{subfigure}
% 		\caption{  In (a), exogenous surfactant in red spreading around a complex section of the maze. We observe corner-hugging behaviour in the top left and bottom branches in this figure. In (b) a COMSOL simulation of $\Gamma_t = \boldsymbol{\nabla}\cdot(\Gamma\boldsymbol{\nabla}\Gamma)/4$ in the same geometry, along with COMSOL's particle tracing feature. Boundary conditions are no-flux of surfactant at every boundary apart from the inlet and outlet with Dirichlet conditions as shown. A small initial concentration profile of half a cosine wave over the length $R$ connects $\Gamma=1$ to $\Gamma= 0.7$ which is the uniform initial condition everywhere else. The position of fluid particles given as red dots, transported from their starting position given by the blue dots, replicates many of the behaviours observed in (a), for example we observe the exogenous surfactant front hugging the inside track around corners, a behaviour that can be understood by observing the compression of level curves around the inside of these bends.  }
% 	\label{fig:2DDynamics}
% \end{figure}

% % \jl{[DISCUSSION/CONCLUSIONS 0.5-0.75 page]}

Our model %\eqref{eq:NonlinearDiff} 
reveals the importance of 
%even a small amount of endogenous surfactant and 
exogenous--endogenous surfactant interaction in the maze-solving behaviour. %Our transport model, taking exogenous and endogenous surfactants as a single concentration field governed by \eqref{eq:NonlinearDiff}, provides further insight. 
Although the  model makes assumptions about the endogenous surfactant, our results show that the model is robust; the maze-solving dynamics continue to be reproduced as the ratio of endogenous to exogenous concentration is varied across one order of magnitude. 
%to these assumptions, reproducing qualitatively, and to some extent quantitatively, 1D and 2D maze-solving dynamics, with possible values for $\delta$ varying across one decade.
%
Through the invisible endogenous surfactant, the exogenous surfactant is `aware' of the whole network topology. 
The large area of the outlet compared to that of the lateral branches ensures that large concentration gradients remain along the solution path, thereby driving the exogenous surfactant to the outlet. In lateral branches, compression of endogenous surfactant limits access by exogenous surfactant. %The combination of asymmetry in the maze network and endogenous surfactant is crucial to the maze-solving behaviour. 
The elliptic nature of the spatial differential operator in \eqref{eq:NonlinearDiff} means that the surfactant spreads through the maze with an \textit{omniscient} view of the geometry, since together the exogenous and endogenous surfactants fill the whole maze at all times. 

Although our model is not designed to simulate surfactant transport in lungs, it may provide  insight and novel methodologies for studying surfactant therapies for lungs. Its essential ingredients, endogenous surfactant \citep{espinosa93,grotberg1995interaction,halpern98,zhang03,stetten2018surfactant,garcia-mouton23} and network asymmetry \citep{horsfield68,tawhai04,schmidt04}, are characteristic of lungs. Exogenous surfactant  delivering drugs to the distal airways and alveoli  may spread non-uniformly through the lung, reducing drug efficacy. The model could inspire lung-scale  models to test the impact of lung network asymmetry on surfactant drug delivery. Our mimetic finite-difference implementation on a network is  amenable to scale-up, as we  found from preliminary tests on a  network with $O(10^4)$ branches based on real lung scans. If the linear regime applies, the transport model  decomposes into a series of modes  computed from the network topology, thus helping understand surfactant transport without computationally expensive simulations. Our model framework  provides a novel approach compared to single-branch models \citep{espinosa93,grotberg1995interaction,halpern98,zhang03,filoche15,copploe19}, which overlook  complex lung topology, and molecular dynamic simulations \citep{laing09,baoukina11,souza18,souza22}, which are numerically  expensive to run over a whole lung.

Our model reveals how the combination of asymmetry in the maze network and exogenous-endogenous surfactant interactions are key to understanding confined transport problems in complex branching networks, such as lungs.

\color{black}
\bigskip
 \begin{acknowledgments}
J.R.L. and O.E.J. acknowledge financial support from EPSRC grant EP/T030739/1. P.L.-F. acknowledges support from NSF CAREER grant 2048234. F.T.-C. acknowledges support from a Distinguished Postdoctoral Fellowship from the Andlinger Center from Energy and the Environment at Princeton University. 
For the purpose of Open Access, the authors have applied a Creative Commons Attribution (CC BY) licence to any Author Accepted Manuscript version arising.
% We thank \dots
 \end{acknowledgments}

% \appendix

% \section{Appendixes}

%\nocite{*}
\newpage 
\bibliographystyle{apsrev4-2}
\bibliography{Refs}% Produces the bibliography via BibTeX.

\end{document}

\thispagestyle{plain}

% title goes here:
\begin{centering}

{ \large \textsc{Supplementary Material} \\ Exogenous--endogenous surfactant interaction yields heterogeneous spreading in complex branching networks
}

\bigskip
% authors go here:

{Richard Mcnair,}\textsuperscript{a}
{Fernando Temprano-Coleto,}\textsuperscript{b}
{Fran{\c{c}}ois J. Peaudecerf,}\textsuperscript{c}
{Fr\'ed\'eric Gibou,}\textsuperscript{d}
{Paolo Luzzatto-Fegiz,}\textsuperscript{d}
{Oliver E. Jensen,}\textsuperscript{a}
{Julien R. Landel}\textsuperscript{a,}\footnote{Corresponding author: \href{mailto:julien.landel@manchester.ac.uk}{julien.landel@manchester.ac.uk}}
\\
\bigskip
{\small
\textsuperscript{a}\textit{Department of Mathematics, University of Manchester, Manchester M13 9PL, UK}
\\
\textsuperscript{b}\textit{Andlinger Center for Energy and the Environment, Princeton University, Princeton, New Jersey 08544, USA}
\\
\textsuperscript{c}\textit{Institut de Physique, Université de Rennes, F-35000 Rennes, France}
\\
\textsuperscript{d}\textit{Department of Mechanical Engineering, University of California, Santa Barbara, California 93106, USA}
\\
}
(Dated: \today)

\end{centering}

\section{Maze geometry and experimental parameters}\label{sec:MazeGeometry}

% \begin{figure}[h]
%     \centering
%      \includegraphics[width=0.4\textwidth]{MazeBranchNumberingScheme.jpg}
%      \caption{Picture at $t^*=\SI{0}{\second}$ of the experiment conducted by Temprano-Coleto \etal \citep{temprano2018soap,temprano2018soap}, showing a numbering scheme for the branches. %The maze is filled with milk and a small quantity of red dye is initially placed at the inlet. At $t^*=\SI{0}{\second}$, a small quantity of surfactant solution is deposited at the inlet. 
%      }
%     \label{fig:numbscheme}
% \end{figure}

\begin{figure}[h]
    \centering
     \includegraphics[width=0.4\textwidth]{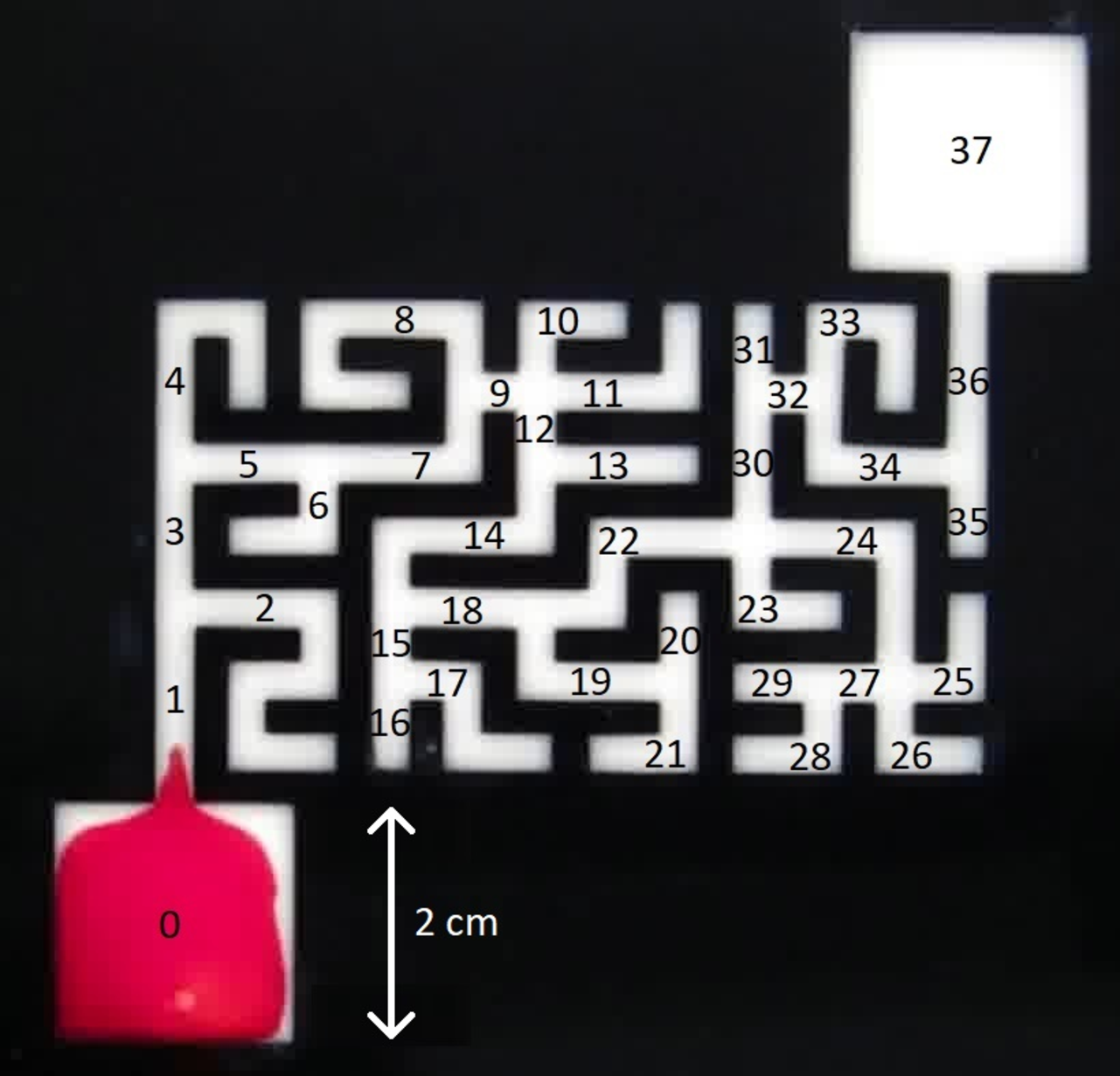}
     ~~~~~~~~~
    \includegraphics[width=0.5\textwidth]{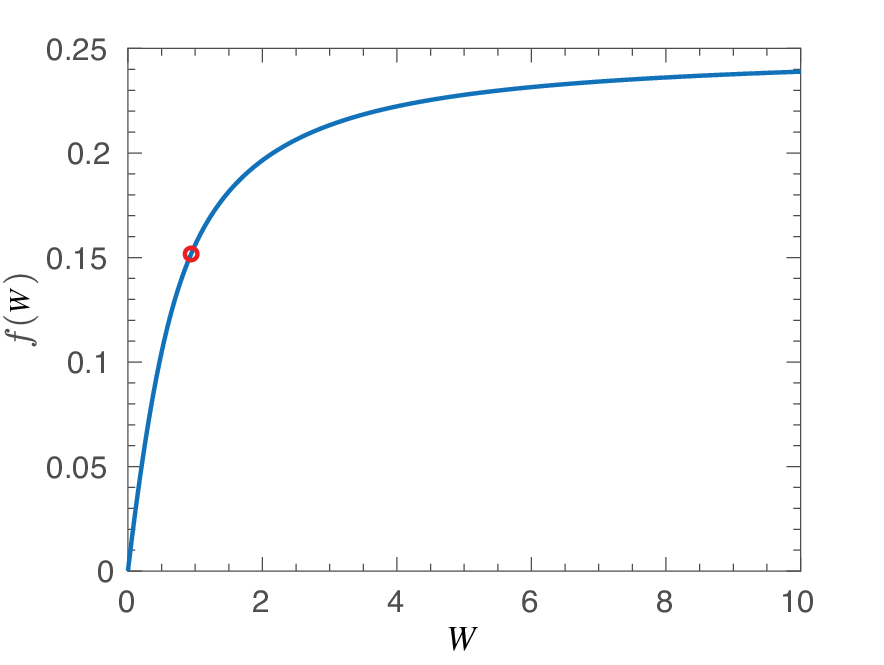}
     \caption{Left: Picture at $t^*=\SI{0}{\second}$ of the experiment conducted by Temprano-Coleto \etal \citep{temprano2018soap}, showing a numbering scheme for the branches. %The maze is filled with milk and a small quantity of red dye is initially placed at the inlet. At $t^*=\SI{0}{\second}$, a small quantity of surfactant solution is deposited at the inlet. 
     Right: Graph of the mobility $f(\AR)$ found from \eqref{FofD}, with the sum truncated at $n=1000$, plotted against the branch aspect ratio $\AR$ in blue, with $f(0.94)$ circled in red. The sum \eqref{FofD} converges to 2 significant figures for $n<100$ for the values of $\AR$ shown. We take $W=0.94$ to be the aspect ratio of the branches of the maze.
     }
    \label{fig:numbscheme}
\end{figure}
% \begin{figure}
%     \centering   
%     \includegraphics[width=0.5\textwidth]{FofD2.eps}
%     \caption{Graph of the mobility $f(\AR)$ found from \eqref{FofD}, with the sum truncated at $n=1000$, plotted against the branch aspect ratio $\AR$ in blue, with $f(0.94)$ circled in red. The sum \eqref{FofD} converges to 2 significant figures for $n<100$ for the values of $\AR$ shown. We take $\AR=0.94$ to be the aspect ratio of the branches of the maze.}
%     \label{fig:FofD}
% \end{figure}

%!!!!I (Richard) have taken out of the folder (and saved to my own computer) all figures not in the current version.!!!

We model the transport dynamics in the  maze experiment conducted by Temprano-Coleto \etal \cite{temprano2018soap}, as presented in their \fig~2 and with the corresponding video showing the large maze at \cite{temprano2018soap} (at 1:47 min). A picture of the experiment at what we define to be the initial time, $t^*=\SI{0}{\second}$ (with a star denoting dimensional quantities), is shown in \fig~\ref{fig:numbscheme} (right), along with a branch numbering scheme. The experimental conditions and protocol are detailed in \cite{temprano2018soap}. 
In table~\ref{tab:expDparameters} we report estimated ranges of values for the material parameters involved in the experiment. For the length of the solution path, the range of values is estimated by measuring the shortest and longest direct line through the solution path. The error associated with the measurement of lengths from image analysis is less than $\SI{1}{\mm}$. The parameter values related to the milk properties, the density $\rho^*$, the dynamic viscosity $\mu^*$ and the nominal surface tension $\gamma_0^*$ (assuming that only natural endogenous surfactant are present at the milk surface), are approximations for the milk-cream mixture used in the experiment \citep{temprano2018soap}. The lowest surface tension of the milk-cream mixture $\gamma_c^*$ when the surfactant concentration is maximum during the experiment is estimated based on a crude approximation for soapy water. The Marangoni force $S^*=\gamma^*_0-\gamma^*_c$, calculated from the values $\gamma^*_0$ and $\gamma^*_c$, gives a sensible range of values.

\begin{table}
    \centering
    \begin{tabular}{|c|c|c|c|} \hline
      Parameter & Symbol & Units & Estimated   \\
      &  &  &  range in  \\    
      &  &  &   experiments  \\    \hline
      Length of solution path & $L_M^*$ & $10^{-1} \ \si{m}$ & $1.36 - 1.84 $\\
      Initial height of liquid film & $h^*_0$ & $10^{-3} \ \si{m}$ & $3.0 -3.5 $\\ 
      Acceleration due to gravity & $g^*$ & $\si{m.s^{-2}}$ & $9.8 $  \\
    Liquid density & $\rho^*$ & $10^3 \ \si{kg.m^{-3}}$ & $0.98-1.0   $\cite{rutz1955some} \\ 
    Dynamic viscosity & $\mu^*$ & $10^{-2} \ \si{Pa.s}$ & $3-4 $ \cite{bakshi1984effect}  \\
    %   Reference surfactant concentration & $\Gamma_c^*$ & $\si{mol.m^{-2}}$ & $10^{-8} - 10^{-5} $\\ 
      Surface tension of milk-cream mixture& $\gamma_0^*$ & $10^{-2} \ \si{N.m^{-1}}$ & $3-5 $\cite{kristensen1997rheology}   \\
      Surface tension of surfactant-laden interface & $\gamma_c^*$ & $10^{-2} \ \si{N.m^{-1}}$ & $2.0-2.5  $ \\ \hline \hline
       Marangoni force & $S^* = \gamma_0^*-\gamma_c^*$ & $ 10^{-2} \ \si{N.m^{-1}}$ & $ 0.5 -3.0$ \\
    \hline
    \end{tabular}
    \caption{Table showing the relevant dimensional parameters of the maze experiment conducted by Temprano-Coleto \etal  \citep{temprano2018soap}. The range of values for the Marangoni force, below the double lines, has been calculated from other values in the table. The parameter values related to the milk properties are approximations for the milk-cream mixture used in the experiment \citep{temprano2018soap}.}
    \label{tab:expDparameters}
\end{table}

\begin{table}
    \centering
    \begin{tabular}{|c|c|c|c|} \hline
     Dimensionless group & Definition & Model  & Estimated   \\  
      &  & assumption & range in experiments  \\  \hline    
      Aspect ratio of liquid film $\epsilon$ & $h^*_0/L_M^*$ & $\ll1$ & $1.7 \times 10^{-2}-2.4 \times 10^{-2}$   \\
      Inverse capillary number $C$ & $\epsilon^2 \gamma^*_c/S^*$ & $\ll1$ & $9.7 \times 10^{-4} - 1.0 \times 10^{-2}$\\
      Bond number $G$ & $\epsilon^2 \rho^* g^* {L_M^*}^2/S^*$ & $\gg1$ & $10 - 40$\\
      Reduced Reynolds number $\epsilon^2 Re$ & $\epsilon^2 \rho^* S^* h_0^*/{\mu^*}^2 $ & $\ll1$ & $ 1.6 \times 10^{-3} - 1.6 \times 10^{-2}$ \\
    \hline
    \end{tabular}
    \caption{Table of relevant dimensionless groups for the maze experiments \citep{temprano2018soap}, calculated using the dimensional parameter values from table \ref{tab:expDparameters}.}
    \label{tab:expNDparameters}
\end{table}

In table~\ref{tab:expNDparameters} we report estimated values for non-dimensional groups. The model we use to simulate the experiment is based on a one-dimensional (1D) lubrication theory approximation to Stokes flow (see section~\ref{sec:TransportEquation}). Inertial forces are dominated by viscous forces when the reduced Reynolds number $\epsilon^2 Re$ is small, where $\epsilon \ll 1$ is the aspect ratio of the main flow path, which is consistent with the values taken from table~\ref{tab:expDparameters}. The equation we use for the transport dynamics assumes that gravity dominates over surface tension and surface tension gradients to maintain a flat surface. Any curvature effects of the film surface associated for instance with the contact angle at the side contact lines are assumed negligible. Accordingly,  the Bond number $G$ is found to be large and the inverse capillary number $C$ small, which is consistent with these assumptions. 
% For simplicity, we neglect any transverse flow dynamics across the width of the branches. As we discuss in section \ref{sec:keyparameters}, this assumption is mostly valid throughout the maze, except at the inlet and outlet square wells and in some side branches.

The dimensionless length $L_i$ of branch $i$, for $i=0,1,2,\dots 37$, is shown in table \ref{tab:BranchLengths}, where we define one unit as the length of the solution path $L_M^*= \SI{155}{mm}$. 
% To model the large square well at the outlet of the maze, we assume that this branch has a much longer length, simulating its larger surface area. The length of the inlet branch is calculated to simulate $M^*_{ex}$ as discussed in section \ref{sec:keyparameters} below. 
These lengths were calculated by counting pixels from the video of the experiment \cite{temprano2018soap}, with all dimensional lengths measured with less than \SI{1}{\mm} accuracy.  The maze video was taken directly overhead, minimizing optical distortion.
% Based on the experimental protocol described in \cite{temprano2018soap}, the inlet and outlet branches $0$ and $37$ in \fig~\ref{fig:numbscheme} have  no-flux boundary conditions imposed at their ends, as do the ends of all the peripheral branches not leading to a junction. 

\begin{table}
    \centering
\begin{tabular}{|c|c|c|c|c|c|c|c|c|c|c|c|c|c|}
\hline
  Branch No. $(i)$   & $0$ & $1$ & $2$ & $3$ & $4$ & $5$ & $6$ & $7$ & $8$ & $9$ & 10 & 11 & 12  \\
  $L_i/10^{-2}$   & $L_0/ 10^{-2}$ & $9.6$ & $19$ & $6.0$ & $14$ & $6.5$ & $6.4$ & $10$ & $16$ & $3.4$ & $6.3$ & $9.6$ & $3.4$  \\
  \hline
     Branch No. $(i)$   & $13$ & $14$ & $15$ & $16$ & $17$ & $18$ & $19$ & $20$ & $21 $& $22$ & $23$ & $24$ & $25$   \\
  $L_i/ 10^{-2}$    & $6.5$ & $13$ & $3.2$ & $3.2$ & $9.9$ & $6.7$ & $6.6$  & $3.2$ & $6.5$ & $13$ & $6.5$ & $13$ & $6.5$  \\
  \hline
  Branch No. $(i)$   & $26$ & $27 $& $28$ & $29$  & $30$ & $31$ & $32$ & $33$ & $34$ & $35$ & $36$ & $37$ &   \\
  $L_i/ 10^{-2}$    & $6.5$ & $3.3$ & $3.2$ & $6.3$  & $6.9$ & $3.1$ & $3.2$ & $9.5$ & $9.9$ & $3.3$ & $7.9$ & $71$  &    \\
  \hline
\end{tabular}
\caption{Table of the branch lengths of the maze, where one unit represents the length of the solution path ($L_M^*=\SI{155}{mm}$). The branch are numbered based on the scheme showed in \fig~\ref{fig:numbscheme} (right).  The length of the inlet branch $L_0$ is defined in section \ref{sec:ContinuumFormulation} as a function of $\delta$, the ratio of endogenous to exogenous initial surfactant concentration.}
\label{tab:BranchLengths}
\end{table}

\section{The surfactant transport equation}
\label{sec:TransportEquation}

The spreading of insoluble surfactant at the free surface of a thin two-dimensional (2D) layer of incompressible viscous fluid under the action of surface-tension gradients can be described by a nonlinear diffusion equation, as we briefly explain.  We use 2D Cartesian coordinates with horizontal coordinate $x^*$. %, %$(x^*,y^*)$, where stars indicate dimensional quantities. We refine our model to the full maze later on. 
A liquid of viscosity $\mu^*$ is confined between a horizontal solid wall %at $y^*=0$ 
and a free surface, at height $h^*(x^*,t^*)$
on which a surfactant with concentration $\Gamma^*(x^*,t^*)$ is present at time $t^*$.  The surface tension $\gamma^*$ of the free surface is assumed to diminish linearly as a function of $\Gamma^*$ for $\delta \Gamma_c^* \leq \Gamma^*\leq \Gamma_c^*$ from a maximum value $\gamma_0^*$ to a minimum value $\gamma_c^*$, with $\delta$ the ratio of endogenous to exogenous initial surfactant concentration such that $\delta \Gamma_c^*$ is the endogenous surfactant concentration initially.  We define $S^*=\gamma_0^*-\gamma_c^*$.  We consider the evolution of $\Gamma^*$ and $h^*$ over horizontal distances $L_M^*$ that are much larger than the characteristic film height $h^*_0$, and define $\epsilon=h^*_0/L_M^*$.  Adopting scales appropriate to lubrication theory, we define $x={x^*}/{L_M^*}$, %$y={y^*}/h^*_0$ 
and $t = {\epsilon S^* t^* }/({L_M^* \mu^*})$. %, with velocity and pressure also scaled appropriately.
%as $u={\mu^*  u^*}/(\epsilon S^*)$ in the longitudinal direction, $v={\mu^*  %v^*}/(\epsilon^2 S^*)$ in the vertical direction and $p = {h^*_0 %p^*}/{S^*}$.  
We define the dimensionless film thickness and surfactant concentration as $h(x,t)=h^*/h^*_0$ and $\Gamma(x,t)=\Gamma^*/\Gamma^*_c$, respectively, and the surface tension as $\gamma = ({\gamma^*-\gamma^*_c})/{S^*}=1-\Gamma$.  Then, when the reduced Reynolds number is sufficiently small ($\epsilon^2 \rho^* S^* h^*_0/{\mu^*}^2 \ll 1$), the flow is governed by the coupled evolution equations \cite{jensen1992insoluble}
\begin{subequations} \label{eq:CoupledEqns}
\begin{align}
h_t - \tfrac{1}{2}\left( h^2 \Gamma_x\right)_x - \tfrac{1}{3}G \left( h^3h_x \right)_x  +\tfrac{1}{3}C \left( h^3 h_{xxx} \right)_x &= 0 , \\
\Gamma_t - \left(\Gamma h \Gamma_x \right)_x - \tfrac{1}{2} G \left( \Gamma h^2 h_x \right)_x + \tfrac{1}{2} C \left( \Gamma h^2 h_{xxx} \right)_x &= 0 ,
\end{align}
\end{subequations}
where $C=\epsilon^2 \gamma_c^*/S^*$ is the inverse capillary number and $G=\epsilon^2 \rho^* g^* {L_M^*}^2/S^*$ the Bond number, which are dimensionless parameters representing the strength of surface tension and gravity relative to surface tension gradients, respectively. Surface diffusion of surfactant and surface rheological stresses are neglected.  Spreading of surfactant from some initial condition can generate deflections of the free surface, although $h$ never exceeds 2 and cannot reach zero in the absence of disjoining pressure, with gravity suppressing deflections at large times  \cite{jensen1992insoluble}. The limit $G\gg 1$ is consistent with the experiment (Table \ref{tab:expNDparameters}). Expanding $h$ and $\Gamma$ in powers of $1/G$ about the base state $h=1$ and imposing no-flux conditions at the ends of the domain, \eqref{eq:CoupledEqns}, to leading order in $\epsilon$, reduces to the nonlinear diffusion equation \cite{jensen1998stress1}
\begin{equation}\label{eq:NonLinDiff1}
	\Gamma_{t} = \tfrac{1}{4}\big(\Gamma \Gamma_{x}\big)_x,
\end{equation} 
which can be rearranged as $\Gamma_{t} = (\Gamma^2)_{xx}/8$.
This captures the essential features of self-induced spreading of an insoluble surfactant on a thin film via Marangoni forces.

\subsection{Surfactant transport along a channel}

The factor of $1/4$ in (\ref{eq:NonLinDiff1}) is appropriate for a thin film, but it takes no account of the lateral no-slip boundary conditions that constrain spreading along a channel in a maze having rectangular cross section with finite width-to-height aspect ratio $\AR$.  We refine the approximation to determine the mobility $f(\AR)$ of the surfactant as follows, where $f\rightarrow 1/4$ in the thin-film limit in which $\AR\rightarrow \infty$.  We again assume (i) no out-of-plane deflection of the air--liquid interface and no externally imposed pressure gradient, so that the volume flux along the channel is uniformly zero, and (ii) the channel is long compared to its width and height so that rapid lateral Marangoni spreading quickly eliminates lateral surfactant concentration gradients \citep{mcnair2022surfactant}.  Under (i) and (ii), the axial velocity component $u^*(x^*,y^*,z^*,t^*)$ is driven by a surfactant gradient $-\Gamma^*_{x^*} (x^*,t^*)$ moving the bulk liquid forward and a counter pressure gradient $-p^*_{x^*}(x^*,t)$ driving a return flow in the bulk (arising via incompressibility from the no-interfacial-deflection condition and the fact that the channels are closed at all ends), where $y^*$ and $z^*$ are coordinates spanning the cross-section of the channel in the  transverse and vertical directions, respectively.  Adopting scalings introduced above, the leading-order axial velocity component satisfies $\nabla_\perp^2 u\equiv u_{yy}+u_{zz}=p_x$ across the cross-section of the channel, with $u=0$ on the bottom and lateral walls of the channel ($z=0$, $y=\pm \AR/2$) and the shear stress condition $u_z=-\Gamma_x$ on the interface at $z=1$.  Linearity allows the field to be decomposed as $u(x,y,z,t)=-p_x(x,t) u^b(y,z)-\Gamma_x(x,t) u^{m} (y,z)$ where $\nabla_{\perp}^2 u^b=-1$ and $u^b_z=0$ on $z=1$, and $\nabla_\perp^2 u^m=0$ and $u^m_z=1$ on $z=1$.  Averaging $u$ over the rectangular cross-section defines volume fluxes $Q^b(\AR)$ and $Q^m(\AR)$, and condition (i) then requires $0=-Q^b p_x -Q^m \Gamma_x$, which determines $p_x$ in terms of $\Gamma_x$.  Finally, the surfactant field is advected by the transversely-averaged surface velocity $\overline{u}_s\equiv \AR{-1} \int_{-\AR/2}^{\AR/2} u(x,y,1,t)\,\mathrm{d}y$, which can be expressed as $\overline{u}_s=-p_x \overline{u}^b_s -\Gamma_x \overline{u}^m_s$.  Eliminating $p_x$ recovers the nonlinear diffusion equation 
\begin{equation}\label{eq:NonLinDiff2}
\Gamma_t = f(\AR) (\Gamma \Gamma_x)_x, \quad \mathrm{where}\quad f(\AR)=\overline{u}_s^m-(\overline{u}_s^b Q^m/Q^b).  
\end{equation}
In the limit of large $\AR$, $u^b=z-\tfrac{1}{2}z^2$, $u^m=z$, $\overline{u}^b_s=\tfrac{1}{2}$, $\overline{u}^m_s=1$, $Q^b=1/3$ and $Q^m=1/2$ giving $f=\tfrac{1}{4}$.

For finite $\AR$, we use separation of variables to obtain the following expressions for $u^b$, $u^m$, $\overline{u}^b_s$, $\overline{u}^m_s$, $Q^b$, $Q^m$ and $f$:
\begin{subequations}
\begin{align}\label{velocA}
u^b(y,z)=& -\tfrac{1}{2}(z^2-2z) \nonumber \\
 -\sum_{n=1}^{\infty} &\frac{2  \sin \left[(n-\tfrac{1}{2}) \pi z\right]\left[ \sinh{\left[(n-\tfrac{1}{2} ) \pi (y+\AR/2)\right]} 
-\sinh\left[(n-\tfrac{1}{2}) \pi (y-\AR/2) \right]  \right]}{\sinh\left[(n-\tfrac{1}{2})\pi \AR\right](n-\tfrac{1}{2})^3\pi^3};\\
u^m(y,z)=& -\sum_{n=1}^{\infty} \frac{\AR(-1)^n \cos{\Big(\frac{2(n-\frac{1}{2}) \pi}{\AR} y\Big)}\sinh{\Big( \frac{2(n-\frac{1}{2}) \pi z}{\AR} \Big)}}{(n-\frac{1}{2})^2\pi^2\cosh{\Big( \frac{2(n-\frac{1}{2}) \pi }{\AR} \Big)}};
% \\
% \overline{u}^b=&
%  -\tfrac{1}{2}(z^2-2z)  -\sum_{n=1}^{\infty} \frac{2\sin{[(n-\frac{1}{2})\pi z ]} [2\cosh{[(n-\frac{1}{2})\pi \AR]}-2]}{\sinh{[(n-\frac{1}{2})\pi \AR]}(n-\frac{1}{2})^4 \pi^4 \AR}, 
 \\
 \overline{u}^b_s=&
 \tfrac{1}{2}+\sum_{n=1}^{\infty} \frac{2(-1)^n [2\cosh{[(n-\frac{1}{2})\pi \AR]}-2]}{\sinh{[(n-\frac{1}{2})\pi \AR]}(n-\frac{1}{2})^4 \pi^4 \AR}; 
 \\
\overline{u}_s^m
=& \sum_{n=1}^{\infty} \frac{\AR \tanh\left[2(n-\frac{1}{2}) \pi  /\AR\right]}{ (n-\frac{1}{2})^3 \pi^3 };
\\
 Q^b=& 1/3 -\sum_{n=1}^{\infty} \frac{4(\cosh{[(n-\frac{1}{2})\pi ]}-1)}{\sinh{[(n-\frac{1}{2})\pi \AR]}(n- \frac{1}{2})^5 \pi^5 \AR} ;
 \\
 Q^m=&  \sum_{n=1}^{\infty} \frac{\AR^2(\cosh{\left[2(n-\frac{1}{2}) \pi/\AR \right]}-1) }{ 2(n-\frac{1}{2})^4 \pi^4  \cosh{\left[2(n-\frac{1}{2}) \pi /\AR\right]} };  \\
\label{FofD}
	f(\AR)=& \sum_{n=1}^{\infty} \frac{\AR \tanh{\Big( \frac{2(n-\frac{1}{2}) \pi }{\AR} \Big)}}{(n-\frac{1}{2})^3\pi^3}  \nonumber
	\\
	- &\frac{\sum_{n=1}^{\infty} \frac{\AR^2(\cosh{\left[2(n-\frac{1}{2}) \pi/\AR \right]}-1) }{ 2(n-\frac{1}{2})^4 \pi^4  \cosh{\left[2(n-\frac{1}{2}) \pi /\AR\right]} }}{ 1/3 -\sum_{n=1}^{\infty} \frac{4(\cosh{[(n-\frac{1}{2})\pi \AR]}-1)}{\sinh{[(n-\frac{1}{2})\pi \AR]}(n- \frac{1}{2})^5 \pi^5 \AR} } 
	\left(\frac{1}{2}+
	\sum_{n=1}^{\infty} \frac{2(-1)^{n} [2\cosh{[(n-\frac{1}{2})\pi \AR]}-2]}{\sinh{[(n-\frac{1}{2})\pi \AR]}(n-\frac{1}{2})^4 \pi^4 \AR}
	\right) . 
\end{align} 
\end{subequations}
Figure~\ref{fig:numbscheme} (left) shows that for $\AR=0.94$, which is the approximate aspect ratio of the channel in the experiment, $f\approx 0.15$, which indicates  a 40\% reduction approximately in the surfactant mobility in comparison to the wide-channel limit. 
% Equation \eqref{eq:NonLinDiff2} describes transport of surfactant along a simple channel; faster transverse spreading, eliminating lateral concentration gradients was addressed in \cite{mcnair2022surfactant}.

In addition to gravitational suppression of interfacial deflections, as used to derive (\ref{eq:NonLinDiff1}) from (\ref{eq:CoupledEqns}), surface tension can also have an significant role in suppressing deflections in a channel flow via pinning of contact lines to the channel edges.  Any interfacial deflection will have two components of curvature: a short length comparable to the channel width; and a longer axial scale induced by competition between viscous and capillary forces, as represented in (\ref{eq:NonLinDiff1}a).  The former can be expected to dominate over the latter, as it acts over shorter lengthscales, and it will act alongside gravity in maintaining a near-uniform liquid depth.  

Transport of a single surfactant species in a shallow channel (\ref{eq:NonLinDiff1}) can be expressed dimensionally as $\Gamma^*_{t^*}+(u^*\Gamma^*)_{x^*}=0$, where $u^*=-\tfrac{1}{4} (h_0^* A^*/\mu^*)\Gamma^*_{x^*}$. Here $A^*=S^*/\Gamma_c^*$ is the activity of the surfactant (the magnitude of the slope of the assumed-linear relationship between surface tension and surfactant concentration).  The equivalent model for a mixture of two surfactant species with concentrations $\Gamma_1^*$ and $\Gamma_2^*$ is $\Gamma^*_{i,t}+(u^*\Gamma_i^*)_{x^*}=0$ for $i=1,2$ with $u^*= -\tfrac{1}{4} (h_0^* /\mu^*)(A_1^*\Gamma^*_{1,x^*}+A_2^*\Gamma^*_{2,x^*})$, again assuming a linear equation of state.  Defining $\Gamma^*=(A_1^*\Gamma_1^*+A_2^*\Gamma_2^*)/(A_1^*+A_2^*)$, $A^*=A_1^*+A_2^*$ and summing the evolution equations weighted respectively by $A_1^*$ and $A_2^*$ recovers exactly the evolution equation for a single species $\Gamma^*$, equivalent to (\ref{eq:NonLinDiff1}).  While for simplicity we assume that endogenous and exogenous surfactant have identical properties in the present analysis, the model accommodates two non-diffusing species with different activities.

% \begin{figure}
%     \centering   
%     \includegraphics[width=0.5\textwidth]{FofD2.eps}
%     \caption{Graph of the mobility $f(\AR)$ found from \eqref{FofD}, with the sum truncated at $n=1000$, plotted against the branch aspect ratio $\AR$ in blue, with $f(0.94)$ circled in red. The sum \eqref{FofD} converges to 2 significant figures for $n<100$ for the values of $\AR$ shown. We take $\AR=0.94$ to be the aspect ratio of the branches of the maze.}
%     \label{fig:FofD}
% \end{figure}

\subsection{Surfactant transport in a maze}
\label{sec:ContinuumFormulation}

We now extend the 1D model to show that exogenous surfactant added to pre-existing endogenous surfactant can simulate the maze-solving behaviour seen in the experiments \cite{temprano2018soap}. This behaviour consists of the exogenous surfactant spreading  from the end of branch $0$ (see numbering scheme in \fig~\ref{fig:numbscheme}, left) to the start of branch $37$, while also not spreading to the end of the lateral branches.

To model the maze-solving behaviour of the out-of-equilibrium surfactant field, we solve the simplified time-dependent surfactant transport equation \eqref{eq:NonLinDiff2} in a network of connected 1D domains (1DDs). Equation \eqref{eq:NonLinDiff2} solves the transport of both the exogenous and endogenous surfactants, as they are assumed to form a single field. Each 1DD  represents one of the $36$ branches of the maze, where the length of the $i$th 1DD, $L_i$, is the length of the $i$th branch. %(made dimensionless using the solution path length $L_M^*= \SI{mm}$, according to the non-dimensional scheme outlined in section \ref{sec:TransportEquation}). 
These lengths are given in table~\ref{tab:BranchLengths}. Additionally, we simulate the square branches $0$ and $37$ at the inlet and outlet of the maze, respectively, with two more 1DDs of longer lengths. The length of the outlet 1DD, $L_{37}$, is calculated to simulate the surface area of branch 37 in the experiment. Counting the number of pixels $P_{37}^*$ constituting branch $37$ in the image in \fig\ref{fig:numbscheme} (left) and, for example $P_1^*$, the number of pixels that constitute branch $1$, we define $L_{37} = L_{1} P_{37}^*/P_{1}^*$. The length of the inlet 1DD simulating branch $0$, $L_0$, is estimated independently to represent the mass of exogenous surfactant $M_{ex}^*$ introduced in this branch at the start of the experiment, as discussed below.

We use subscripts to number the 1DDs, so that for example $\Gamma_0(x_0,t)$ denotes the concentration in the inlet 1DD, for $0\leq x_0 \leq L_0$ along the branch, and at time $t$. Each 1DD receives a coordinate direction such that, for example, along 1DD $i$ the coordinate direction is $x_i$, and this coordinate increases in the direction away from the junction nearest to the inlet 1DD. 
We solve the governing equation \eqref{eq:NonLinDiff2} for the surfactant transport equation throughout the 1D network along with the following initial, continuity  and boundary conditions, which simulate the conditions in the maze experiment,
\begin{equation}\label{eq:initcond1}
    \Gamma_0(x_0,0) = 1 \quad \text{for all } 0\leq x_0 \leq L_0,
\end{equation}
\begin{equation}\label{eq:initcond2}
    \Gamma_1(x_1,0) = 
    \begin{cases}
    \frac{1}{2}(1-\delta)\cos{(\pi x_1/x_f)} + \frac{1}{2}(1+\delta) \quad &0\leq x_1\leq x_f ,\\
    \delta \quad &x_f < x_1 \leq L_1,
    \end{cases} 
\end{equation}
\begin{equation}\label{eq:initcond3}
 \Gamma_i(x_i,0) = \delta \quad \text{for all } 0\leq x_i \leq L_i, \quad \text{for $i=2,3, \dots, 37$}.
\end{equation}
\begin{equation}\label{eq:noflux1}
 \frac{\partial}{\partial x_0} \Gamma_0(0,t) = 0 , \quad \frac{\partial}{\partial x_{37}} \Gamma_{37}(L_{37},t) = 0, 
\end{equation}
\begin{equation}\label{eq:noflux2}
 \frac{\partial}{\partial x_i} \Gamma_i(L_i,t) = 0 
\end{equation} 
for all 1DDs $i$ not connecting to a junction at $ L_i$,
\begin{equation}\label{eq:ConcCont}
    \Gamma_j(0,t) = \Gamma_i(L_i,t)
\end{equation} 
for every 1DD $i$ terminating at a junction, and for all 1DDs $j \in J(i)$ where $J(i)$ is the set of all branches such that $x_j = 0$ locates the same junction as $x_i = L_i$. Finally,
\begin{equation}\label{eq:FluxCond}
     \frac{\partial}{\partial x_i} \Gamma_i(L_i,t) = \sum_{j \in J(i)} \frac{\partial}{\partial x_j} \Gamma_j(0,t) ,
\end{equation} 
for the same set of 1DDs as \eqref{eq:ConcCont}. 
The initial, continuity and boundary conditions stated above, \eqref{eq:initcond1}--\eqref{eq:FluxCond}, closely model the experimental conditions described in section \ref{sec:MazeGeometry}. The initial condition in the 1DD representing the inlet branch, $i=0$, is that the surfactant concentration is uniformly equal to $1$ \eqref{eq:initcond1}. All the other 1DDs  have a uniform surfactant concentration of smaller value $\delta <1$, which represents the initial ratio of concentrations of endogenous surfactant to exogenous surfactant \eqref{eq:initcond3}. The only exception is in the 1DD which connects the inlet 1DD to the rest of the maze, where a short section of length $x_f=0.032$ smoothly connects the two uniform surfactant concentrations with half a cosine wave \eqref{eq:initcond2}. The dynamics are driven by the gradient of the surfactant concentration, so we impose the initial condition to be $C^1$ continuous so that the initial velocity is defined everywhere. The smooth profile in branch 1 has been added to avoid numerical issues at early times. Its influence on the overall transport dynamics is negligible.
The boundary condition at unconnected ends of 1DDs in the network is no flux of surfactant, such as at the start of the inlet and end of the outlet branches \eqref{eq:noflux1}, and at the end of the peripheral 1DDs \eqref{eq:noflux2}. At junctions where multiple 1DDs are connected, the conditions are continuity of surfactant concentration \eqref{eq:ConcCont}, and continuity of flux of surfactant \eqref{eq:FluxCond}.

\subsection{Boundary between the exogenous and endogenous surfactants}

As shown in \cite{grotberg1995interaction}, in the absence of surface diffusion, the front of exogenous surfactant, added to a surface with a pre-existing endogenous surfactant concentration, moves like a material element transported at the surface velocity generated through the surfactant-induced Marangoni stress.  To simulate the experiment, the evolution of the front location $x_j=\widetilde{\Lambda}_j(t)$ (the tilde denotes a simulated solution to avoid confusion with the experimental data $\Lambda_j$) of exogenous surfactant in 1DD $j$, given (\ref{eq:NonLinDiff2}), satisfies
\begin{equation}\label{eq:FrontLocation}
  \frac{\mathrm d \widetilde{\Lambda}_j(t)}{\mathrm d t} = 
 - f(\AR) \frac{\partial}{\partial x_j}\Gamma_{}(x_j=\widetilde{\Lambda}_j(t),t),
\end{equation} 
for all 1DDs $j$ where a front exists. When $\Lambda_j(t) = L_j$, the front ceases to exist in 1DD $j$, and appears in all 1DDs $i\in J(j)$, where $J(j)$ is the set of 1DDs originating at the head junction of 1DD $j$. The initial conditions are that initially a front  only exists in 1DD $1$, and
\begin{equation}\label{eq:FrontInitCond}
    \widetilde{\Lambda}_1(0) = x_f,
\end{equation} 
and we impose that $ \widetilde{\Lambda}_i(t_b) = 0 $ whenever the front in 1DD $i$ appears at time $t =t_b$. The equation (\ref{eq:FrontLocation}) is solved simultaneously with \eqref{eq:NonLinDiff2}. 
% When a front enters a 1DD $i$ not connected at its downstream junction to any other 1DDs, \hbox{i.e.} where the no-flux boundary condition \eqref{eq:noflux2} applies at $x_i=L_i$, the front will naturally stop at some $x_i<L_i$, at large time, owing to the presence of endogenous surfactant occupying $\widetilde{\Lambda}_{i}\leq x_i\leq L_i$. We also expect receding behaviour in some of the peripheral 1DDs nearest to the inlet (as observed in the experiments), as the concentration rises and then falls in the first few 1DDs of the solution path. 
The maze simulation is completed when $\widetilde{\Lambda}_{36} = L_{36}$.

\subsection{Key non-dimensional parameters in the model}\label{sec:keyparameters}

The model has three unknown parameters to be determined empirically. The first is $\delta$, the ratio of initial endogenous surfactant concentration to the reference exogenous surfactant concentration. The second is $\tau_{end}$, which is the ratio of simulation completion time to experiment completion time. We use $\tau_{end}$ to improve the comparison to the experiment of the dynamic behaviour of exogenous surfactant along the solution path (see section~\ref{sec:ErrorComparison} for further details), but $\tau_{end}$ does not affect the model dynamics significantly. The third \textit{a priori} unknown parameter is $L_0$, the length of the inlet 1DD. As detailed below, the length $L_0$ is used to simulate the mass of exogenous surfactant $M_{ex}^*$ in the inlet branch 0.

Examining the video of the experiment \cite{temprano2018soap}, we observe complex dynamics occurring in the inlet branch $0$ at early times. The exogenous surfactant is not added instantaneously, nor is it added to the entire area initially occupied by the dye. Part of branch $0$ is unoccupied by the dye, and remains unoccupied throughout the duration of the experiment. Moreover, our quantitative analysis of how the dye front moves at early times from branch 0 to branch 1 shows a very fast dynamics, with a short time scale characteristically different from the dynamics throughout the internal branches of the maze. We assume that this rapid dynamics is associated with the inlet branch being a large square, rather than a narrow channel, such that the dynamics is mostly 2D, rather than 1D as assumed in our model. To avoid adding unnecessary complexity to our model, we do not model  this early time  dynamics at the inlet branch as it has limited impact on the long-time dynamics throughout the maze. We find that the overall maze dynamics is well-captured by a 1D inlet branch, whose length  is slightly different from the geometric value. Nevertheless, we still assume that the inlet branch is occupied uniformly by exogenous surfactant at $t=0$. Thus, the inlet branch length is chosen to represent the mass of exogenous surfactant deposited at $t=0$, rather than its geometric size. The actual size of branch $0$ is not crucial to explaining the key behaviour of the experiment (unlike branch $37$). The role of branch $0$ is to act as a (finite) reservoir of exogenous surfactant. 

Owing to conservation of mass, we can estimate the exogenous mass of surfactant $M_{ex}^*$, which is used to determine $L_0$, by examining the video of the experiment at late times. 
% We obtain an approximation for the ratio of exogenous to endogenous surfactant mass that was present in the experiment. 
At late times, the surface tension throughout the maze approaches a constant value. Under the modelling assumption that the material parameters of the exogenous surfactant are the same as the endogenous surfactant, this uniform surface tension means that the concentrations of both species of surfactant also evolve to a constant value, which we call $\Gamma_u^*$.
Furthermore, under the assumptions that the surfactant species do not mix and that the red dye closely follows the exogenous--endogenous interface, the endogenous surfactant occupies the area of the maze which is white, and the exogenous surfactant occupies the area of the maze which is red. Calling the dimensional masses of exogenous and endogenous surfactant $M_{ex}^*$ and $M_{en}^*$ respectively, we make the approximations that
\begin{equation}
    M_{en}^*  \approx \int_{A_{w}^{\infty *}} \Gamma_{u}^* \ \mathrm{d}S^* = \Gamma_u^* A_{w}^{\infty *}, \quad \text{and} \quad M_{ex}^*  \approx \int_{A_{r}^{\infty *}} \Gamma_{u}^* \ \mathrm{d}S^* = \Gamma_u^* A_{r}^{\infty *},  
    \end{equation}
% and 
%     \begin{equation}
%     M_{ex}^*  \approx \int_{A_{r}^{\infty *}} \Gamma_{u}^* \ \mathrm{d}S^* = \Gamma_u^* A_{r}^{\infty *},
%     \end{equation}
where $A_{w}^{\infty *}$ and $A_{r}^{\infty *}$ are the surface areas of the maze not occupied by the dye, and occupied by the dye respectively when the system reaches a steady state at the end of the video. The concentration $\Gamma_u^*$ cancels when calculating the ratio of masses, which we exploit to impose the condition that the ratio of dimensionless masses in our model to be the same as the ratio of masses in the experiment. The non-dimensional exogenous surfactant mass relates to the dimensional mass by $M_{ex}^* = M_{ex}\Gamma_c^* L_M^* W^*$, where $W^*$ is the dimensional width of the maze branches, with a similar identity holding for the endogenous surfactant mass $M_{en}$. Thus, $M_{ex}/M_{en} = A_{r}^{\infty *}/A_{w}^{\infty *}$ provides an identity for $M_{ex}$ in terms of $\delta$
    \begin{equation}\label{eq:Mex1}
        M_{ex} = M_{en} \frac{  A_{r}^{\infty *}}{ A_{w}^{\infty *}} = \delta  \left[ \sum_{i=1}^{N} L_i - x_f \right]  \frac{  A_{r}^{\infty *}}{ A_{w}^{\infty *}},
    \end{equation} 
where the quantity $\left[ \sum_{i=1}^{N} L_i - x_f \right]$ is the total length of all the 1DDs in the model not initially occupied by the exogenous surfactant.
From the initial conditions, the exogenous surfactant mass satisfies
\begin{multline}\label{eq:Mex2}
    M_{ex} = \int_{0}^{L_0} \Gamma_0(x_0,0) \mathrm{d}x_0 + 
    \int_{0}^{x_f} \Gamma_1(x_1,0) \mathrm{d}x_1
    \\
    = \int_{0}^{L_0} 1 \ \mathrm{d}x_0 + \int_{0}^{x_f} \left[ \frac{1}{2}(1-\delta)\cos{(\pi x_1/x_f)} + \frac{1}{2}(1+\delta) \right] \ \mathrm{d}x_1 = L_0 + \frac{1+\delta}{2}x_f.
\end{multline} 
Using \eqref{eq:Mex1} and \eqref{eq:Mex2}, the dimensionless length of the inlet 1DD ($i=0$) can be estimated as
\begin{equation}\label{eq:L_0}
    L_0 =  \delta  \left[ \sum_{i=1}^{N} L_i - x_f \right]  \ \frac{  A_{r}^{\infty *}}{ A_{w}^{\infty *}}  - \frac{1+\delta}{2}x_f,
\end{equation}
where $A_{r}^{\infty *}$ and $A_{w}^{\infty *}$, are measured from the video, the dimensionless lengths $L_i$, $i=1,\dots,N$ are taken from table~\ref{tab:BranchLengths}. %, and $x_f$ is calculated from the position of the exogenous front in  \figref{fig:numbscheme} (left), which is the experiment at $t=0$. 
We note that the precise value of  $x_f$ (chosen as 0.032) has negligible impact on the maze dynamics. The remaining dimensionless parameters $\delta$ and $\tau_{end}$ are determined through comparison with the experiment (see section~\ref{sec:ErrorComparison}).

\section{Numerical methods}
\label{sec:MFD}

\subsection{Spatial discretisation}

We model the transport dynamics in the maze by the time-dependent equation (\ref{eq:NonLinDiff2}) along a connected network of 1DDs, subject to the  initial, continuity and boundary conditions \eqref{eq:initcond1}--\eqref{eq:FluxCond} presented in section~\ref{sec:ContinuumFormulation}. The endogenous--exogenous surfactant interface, assumed to be the front of the red dye, is calculated through \eqref{eq:FrontLocation} under the initial condition \eqref{eq:FrontInitCond}.
We solve a discretised approximation of the continuum formulation of the problem on a network having the same topology as the maze, exploiting the methodology of mimetic finite differences.
We represent the maze as a directed graph. Every junction and every unconnected end of a 1DD is represented by a vertex, and within each 1DD we add vertices to refine the discretisation with oriented edges between them. The number of vertices $\mathbb{N}_i$ within the $i$th 1DD is chosen such that these vertices are equally spaced at a distance $\Delta x_i = L_i/(\mathbb{N}_i+1)$, where $\Delta x_i$ is as close as possible to a chosen global discretisation value $\Delta x$ such that it gives an integer $\mathbb{N}_i$. This is accomplished by
\begin{equation}\label{eq:Deltax_i}
    \Delta x_i = \frac{L_i }{ \text{round}(L_i/\Delta x)}.
\end{equation} 
As a convention, we choose the orientation of every edge to point away from the vertex nearest to the inlet. Following the orientation of each edge, its vertex nearest to the inlet is designated as tail vertex, and its other vertex is designated as head vertex.
The topology of the directed graph is encapsulated by its signed incidence matrix which is defined by
\begin{equation}\label{Incidence}
	A_{ij} = \begin{cases}
		+1 \ \ \ \ \ &\text{Edge } i \sim \text{Vertex } j | \text{ Edge $i$ points towards Vertex $j$}, \\
		-1 \ \ \ \ \ &\text{Edge } i \sim \text{Vertex } j | \text{ Edge $i$ points away from Vertex $j$}, \\
		0 \ \ \ \ \ &\text{Edge } i \nsim \text{Vertex } j, \\
	\end{cases}
\end{equation} 
with $i$ and $j$ two integers such that $1\leq i \leq m$, $1\leq j \leq n$ for a graph with $m$ edges and $n$ vertices, and where $\sim$ and $\nsim$ mean `connected to' and `not connected to', respectively.  Each row of $\mathsf{A}$ corresponds to an edge of the graph and has exactly two non-zero entries: $+1$ corresponding to the head vertex of the  edge, and $-1$ corresponding to the  tail vertex of the edge.  We designate $\mathsf{A}_u$ the unsigned incidence matrix, which is the defined via $A_{u(ij)} = |A_{ij}|$.  The labelling of edges and vertices is arbitrary.  A subset of adjacent vertices is represented by a vector (or chain) $\mathbf{D}$ of length $n$ whose components are $1$ for every vertex in the subset and zero otherwise.  Non-zero components of the chain $\mathsf{A}\mathbf{D}$ identify the oriented edges bounding the domain represented by $\mathbf{D}$.

To define physically meaningful operators on the graph, we include metric information describing the physical system represented by the graph. We define the edge length metric tensor as the diagonal matrix of size $m\times m$
\begin{equation}
\label{eq:edgemetric}
    \mathsf{L} = \text{diag}( \Delta x_0, \dots, \Delta x_0, \dots, \Delta x_1, \dots, \Delta x_1, \dots \Delta x_2, \dots),
\end{equation} where $\Delta x_0$, $\Delta x_1$, etc. are the lengths of the edges of the graph, each obtained from \eqref{eq:Deltax_i}, ordered the same as the rows of $\mathsf{A}$.  We define the associated vertex length metric tensor as
\begin{equation}
    \mathsf{V}=\mathrm{diag}\left(\tfrac{1}{2}\mathsf{A}_u^T \mathsf{L}\mathbb{1}_e \right),
\end{equation}
where $\mathbb{1}_e$ is the $m$-vector (or chain) $(1,1,\dots,1)$ identifying all the edges of the graph. The quantity $V_{jj}$ is half the sum of edge lengths of all edges connected to vertex $j$.  
We define $\boldsymbol{\Gamma}_v$ as the $n$-vector (co-chain) of surfactant concentrations defined on the $n$ vertices of the graph, and $\boldsymbol{\Gamma}_v^2$ as the component-wise square of this vector.  The incidence matrix can be used as an operator which can act on $\boldsymbol{\Gamma}_v^2$.  For instance,
%\begin{equation}
$\mathsf{A}\boldsymbol{\Gamma}_v^2$
%\end{equation} 
is a differencing operation on $\boldsymbol{\Gamma}_v^2$ that returns a co-chain of variables defined on the edges of the graph, where the value on each edge is the difference of the square of the concentration on the bounding vertices of the edge. Thus, using (\ref{eq:edgemetric}) we can  define a gradient operator on $\boldsymbol{\Gamma}_v^2$ as $
    \mathsf{L}^{-1}\mathsf{A}\boldsymbol{\Gamma}_v^2$. This is equivalent to %the differencing operation, but each edge variable is now divided by the length of the edge, and so is 
a second-order-accurate central-difference approximation of the gradient of $\boldsymbol{\Gamma}_v^2$ defined at the midpoint of each edge. 
%We solve equation 
%\begin{equation}
% \frac{\dee \Gamma}{\dt} = \frac{1}{4}{\boldsymbol \nabla} \cdot ( \Gamma {\boldsymbol \nabla} \Gamma ),
%\end{equation}
%by representing our maze as a directed graph.
%\begin{equation}\label{Incidence}
%	\mathsf{A}_{ij} = \begin{cases}
%		+1 \ \ \ \ \ &\text{Edge } i \sim \text{Vertex } j | \text{ Edge $i$ points towards Vertex $j$}, \\
%		-1 \ \ \ \ \ &\text{Edge } i \sim \text{Vertex } j | \text{ Edge $i$ points away from Vertex $j$}, \\
%		0 \ \ \ \ \ &\text{Edge } i \nsim \text{Vertex } j. \\
%	\end{cases}
% \end{equation} where $\sim$ means `connected to'. 
We can therefore approximate the continuous flux $-f(\AR)(\Gamma^2)_x/2$ in the rearranged version of (\ref{eq:NonLinDiff2}) by the discrete expression
\begin{equation}
    \mathbf{q}_e = -\frac{f(\AR)}{2}  \mathsf{L}^{-1}\mathsf{A}\boldsymbol{\Gamma}_v^2.
\end{equation} The flux $\mathbf{q}_e$ is defined on the edges of the graph. 
The vector $\mathsf{V}\boldsymbol{\Gamma}_v$ can be interpreted as the mass of sufactant associated with each vertex.  Then $\langle \mathbf{D},\boldsymbol{\Gamma}_v\rangle\equiv \mathbf{D}^T\mathsf{V}\boldsymbol{\Gamma}_v$ can be interpreted as the mass of surfactant in the simply connected domain represented by the chain $\mathbf{D}$.  Mass conservation at each individual vertex (in the absence of sources and sinks) is expressed as 
\begin{equation}
\label{eq:mass}
    \frac{\mathrm{d}}{\mathrm{d}t} \mathsf{V}\boldsymbol{\Gamma}_v=\mathsf{A}^T\mathbf{q}_e.
\end{equation}
Here, $\mathsf{A}^T\mathbf{q}_e$ returns a co-chain of variables defined on the vertices of the graph giving the net amount of flux entering each vertex. Indeed, each row of $\mathsf{A}^T$ corresponds to a vertex of the graph, where the components in the row are either $+1$ or $-1$ for each edge (corresponding to its column) pointing towards or pointing away from that vertex, respectively.  Equivalently, from (\ref{eq:mass}), $-\mathsf{V}^{-1} \mathsf{A}^T$ is the analogue of the divergence operator applied to all the vertices. Therefore, for a fixed metric $\mathsf{V}$ in time, the discrete analogue of the surfactant transport equation \eqref{eq:NonLinDiff2} in a maze of a given topology is
\begin{equation} \label{SystemOdesFlux1}
    \frac{\mathrm{d }\boldsymbol{\Gamma}_v}{\mathrm{d} t} = \mathsf{V}^{-1}\mathsf{A}^T \mathbf{q}_e= -\frac{f(\AR)}{2} \mathsf{V}^{-1}\mathsf{A}^T   \mathsf{L}^{-1}\mathsf{A}\boldsymbol{\Gamma}_v^2.
 \end{equation} 
Left-multiplying (\ref{SystemOdesFlux1}) by $\mathbf{D}^T\mathsf{V}$ gives
 \begin{equation}
     \frac{\mathrm{d}}{\mathrm{d}t} \langle \mathbf{D}, \boldsymbol{\Gamma}_v\rangle = (\mathsf{A}\mathbf{D})^T \mathbf{q}_e.
 \end{equation}
 We recognise $(\mathsf{A}\mathbf{D})^T \mathbf{q}_e$ as the net flux entering the boundary of the domain represented by $\mathbf{D}$.  This provides 
% If we define a subset of the vertices of our graph as a chain $\mathbf{D}$ whose components are $1$ for every component we include in our subset, and $0$ otherwise, then the mass of surfactant in this subset is given by the discrete integral $\mathbf{D}^T\mathsf{V} \boldsymbol{\Gamma}_v$, so if we left multiply \ref{SystemOdesFlux1} by $\mathbf{D}^T\mathsf{V}$ we get
%\begin{equation} \label{SystemOdesFlux2}
%    \frac{\partial (\mathbf{D}^T\mathsf{V} \boldsymbol{\Gamma}_v)}{\partial t} = - \mathbf{D}^T\mathsf{V}\mathsf{V}^{-1}\mathsf{A}^T \mathbf{q}_e = -\mathbf{D}^T \mathsf{A}^T \mathbf{q}_e
%\end{equation} 
%$- \mathbf{D}^T \mathsf{A}^T = -(\mathsf{A} \mathbf{D} )^T$ which is equal to the transpose of a vector of edge variable which is equal to $-1$ where an edge leaves our subset of vertices, and $+1$ where an edge enters into our subset ($\mathsf{A}\mathbf{D}$ is the differencing operator acting on $\mathbf{D}$). The quantity $-\mathbf{D}^T \mathsf{A}^T \mathbf{q}_e$ is therefore a discrete integral of the flux over the bounding edges of the subset of vertices defined by $\mathbf{D}$, and so \eqref{SystemOdesFlux2} says that the change in concentration integrated over our subset is equal to the integral of the fluxes defined over the boundary of our subset, and is thus 
a discrete analogue of the divergence theorem on the graph.  When implemented numerically, this formulation conserves mass to machine precision.

% \section{Convergence of the numerical scheme}

% To discretize  a physical maze, we  construct a graph representing the topology of the maze, where every channel end and every junction is represented by a vertex. For a given discretization value $\Delta x$ of the maze, we find the closest value to $\Delta x$ which divides each branch length \jl{so as to give an integer $N_b$ for this branch}. We then add $N_b$ vertices equally spaced along the branch $b$, giving a higher spatial resolution to the discretized maze. An edge is defined as a section of a branch connecting two vertices, and is assigned an orientation from the tail vertex to the head vertex. We choose  the tail vertex to be the vertex nearest to the inlet. This arbitrary choice does not affect the transport dynamics. 

\subsection{Temporal discretisation}

We solve \eqref{SystemOdesFlux1} using a semi-implicit time-stepping finite-difference scheme. At the $(k+1)$th time step, we approximate $\boldsymbol{\Gamma}_v^2$ as 
 $ \mathsf{Q}\boldsymbol{\Gamma}_v^{(k+1)}$ 
where $\mathsf{Q} = \text{diag}(\boldsymbol{\Gamma}_v^{(k)})$. Therefore, (\ref{SystemOdesFlux1}) can be approximated by
\begin{equation}
    \frac{\boldsymbol{\Gamma}_v^{(k+1)} - \boldsymbol{\Gamma}_v^{(k)} }{\Delta t}= -\frac{f(\AR)}{2} \mathsf{V}^{-1}\mathsf{A}^T   \mathsf{L}^{-1}\mathsf{A}\mathsf{Q}\boldsymbol{\Gamma}_v^{(k+1)}.
    \end{equation} The solution at the $(k+1)$th time step is then obtained by solving the linear system
\begin{equation}\label{eq:NumericalScheme}
     \left( \mathsf{I} + \frac{f(\AR)}{2}{\Delta t}\mathsf{V}^{-1}\mathsf{A}^T   \mathsf{L}^{-1}\mathsf{A} \mathsf{Q} \right)  \boldsymbol{\Gamma}_v^{(k+1)} = \boldsymbol{\Gamma}_v^{(k)} ,
\end{equation} where $\mathsf{I}$ is the $n\times n$ identity matrix.
As discussed, the boundary conditions \eqref{eq:noflux2} to \eqref{eq:FluxCond} are implemented automatically within the formulation \eqref{eq:NumericalScheme}. The initial conditions are imposed by projecting \eqref{eq:initcond1}--\eqref{eq:initcond3} onto the vertices of the graph, whereby we obtain the initial concentration vector $\boldsymbol{\Gamma}_v^{(0)}$. This is done using the same order for the components of $\boldsymbol{\Gamma}_v^{(0)}$ as we have ordered the columns of $\mathsf{A}$ (this ordering is arbitrary but must be consistent).
The numerical scheme is  found to converge with second-order spatial accuracy following $\Delta x^2$, with $\Delta x\to 0$.  This is consistent with Brio et al. \cite{brio2022spectral}, who found that the rows of (\ref{SystemOdesFlux1}) are a second-order accurate approximation of the Laplacian at each vertex of a graph.

Equation \eqref{eq:FrontLocation} is solved simultaneously with \eqref{eq:NumericalScheme} to keep track of the fronts of exogenous surfactant. We define $\boldsymbol{\Lambda}^{(k)}$ as the vector of locations of exogenous surfactant fronts. This vector has the same number of components as the number of edges in the graph. 
To impose the initial conditions \eqref{eq:FrontInitCond}, every component of $\boldsymbol{\Lambda}^{(0)}$ is zero except the $\lfloor x_f/\Delta x_1\rfloor$ components which represent the first $\lfloor x_f/\Delta x_1 \rfloor$ edges in 1DD $1$, where $\lfloor  a  \rfloor$ means the floor of $a$, returning the largest integer smaller than $a$. These components of $\boldsymbol{\Lambda}^{(0)}$ are set to $\Delta x_1$. We set the $\lfloor x_f/\Delta x_1\rfloor +1 $ component in 1DD $1$ equal to $x_f-\Delta x_1\lfloor x_f/\Delta x_1 \rfloor$.

We also equip our scheme with an indicator vector $\boldsymbol{I}^{(k)}$ which has  the same number of components as there are edges. Each component is either $1$ if a front of the exogenous surfactant is present in that edge or $0$ if not. We set the $\lfloor x_f/\Delta x_1 \rfloor+1$ component of $\boldsymbol{I}^{(0)}$ in 1DD $1$ to $1$, and every other component to $0$. The numerical scheme which solves equation \eqref{eq:FrontLocation}, relating time-step $k$ to $k-1$, is
\begin{equation}\label{eq:NumericalFrontLocation}
    \boldsymbol{\Lambda}^{(k)} = \boldsymbol{\Lambda}^{(k-1)} - f(\AR)  \boldsymbol{I}^{(k)} \circ \mathsf{L}^{-1} \mathsf{A} \boldsymbol{\Gamma}_v^{(k-1)} ,
\end{equation}
where $\circ$ is the component-wise product of vectors. 

The $j$th component of $\boldsymbol{I}^{(k-1)} \circ \boldsymbol{\Lambda}^{(k)}$, if it is non-zero, corresponds to an exogenous surfactant front within the $j$th edge of the graph. After time-step $k$ we check whether every non-zero component of $\boldsymbol{I}^{(k-1)} \circ \boldsymbol{\Lambda}^{(k)}$ is within some  small tolerance $\varepsilon$ of $\Delta x_i$, where $i$ is the index of the 1DD where the edge corresponding to the component is located. If $|\Delta x_i - \boldsymbol{\Lambda}^{(k)}_j| < \varepsilon$ for this component $j$, we `hand over' this surfactant front to the connecting edges by setting the component of $\boldsymbol{I}^{(k)}$ for that edge to $0$, and the forward connecting edges to $1$. The other components of $\boldsymbol{I}^{(k)}$ remain the same as $\boldsymbol{I}^{(k-1)}$. 
To reduce error we also set the components corresponding to the forward connecting edges of $\boldsymbol{\Lambda}^{(k)}$ to $\Delta x_i - \boldsymbol{\Lambda}^{(k)}_j$.
We also take into account receding effects by testing whether or not the fronts have receded to the start of any edge, in which case we pass the front location back to the previous edge following a similar algorithm.
The location of the exogenous surfactant front within any 1DD at time-step $k$ is given by the sum of the components of $\boldsymbol{\Lambda}^{(k)}$ corresponding to the edges representing that 1DD. If we call $J(i)$ the set of indices corresponding to the edges discretising 1DD $i$, and we assume a uniform time step $\Delta t$,
\begin{equation}\label{eq:LeadingEdgeSolution}
    \widetilde{\Lambda}_i(k\Delta t) = \sum_{j \in J(i)} \boldsymbol{\Lambda}^{(k)}(j).
\end{equation}

\section{Linearisation and modal decomposition of the transport model}
% \section{Modal structure of the transport operator}

\subsection{Linear modes}

We can approximate the nonlinear diffusion equation \eqref{eq:NonLinDiff2} in the limit of small gradients of surfactant concentration as a linear diffusion equation. If we assume that the surfactant concentration has the form $\Gamma = \bar{\Gamma}+ \hat{\Gamma}(x,t)$ where $\hat{\Gamma}\ll \bar{\Gamma}$, the nonlinear diffusion equation projected onto the discrete representation of the maze (\ref{SystemOdesFlux1}) becomes, to leading order, 
 \begin{equation}
 \label{eq:lindiff}
     \frac{\mathrm{d}\hat{\boldsymbol{\Gamma}}}{\mathrm{d}t} = -\kappa\mathsf{V}^{-1}\mathsf{A}^T\mathsf{L}^{-1} \mathsf{A} \hat{\boldsymbol{\Gamma}},
 \end{equation} 
where $\kappa = f(\AR)\bar{\Gamma}$, and we identify the Laplacian operator
\begin{equation}\label{eq:VertexLaplacian}
    \kappa\mathsf{V}^{-1}\mathsf{A}^T\mathsf{L}^{-1} \mathsf{A}.
\end{equation}
If we define $\mathbf{U} = \mathsf{V}^{1/2}\hat{\boldsymbol{\Gamma}}$ then (\ref{eq:lindiff}) becomes
  \begin{equation}
     \mathbf{U}_t = -\kappa\mathsf{V}^{-1/2}\mathsf{A}^T\mathsf{L}^{-1} \mathsf{A} \mathsf{V}^{-1/2} \mathbf{U}.
\end{equation} 
The operator $\kappa \mathsf{V}^{-1/2}\mathsf{A}^T\mathsf{L}^{-1} \mathsf{A} \mathsf{V}^{-1/2}$ is symmetric, and therefore has orthogonal eigenvectors, so that $\mathbf{U}_i^T\mathbf{U}_j = \delta_{ij}$ for normalised eigenvectors $\mathbf{U}_i$ and $\mathbf{U}_j$, with distinct eigenvalues $\lambda_i$ and $\lambda_j$ (as the maze is not symmetric, all eigenvalues are distinct), and
where $\delta_{ij}$ is the Kronecker delta. Thus, we can express $\hat{\boldsymbol{\Gamma}}$ in terms of orthonormal eigenvectors $\boldsymbol{\phi}_i = \mathsf{V}^{-1/2}\mathbf{U}_i$ of \eqref{eq:VertexLaplacian} that satisfy 
\begin{equation}
     \boldsymbol{\phi}_i^T\mathsf{V}\boldsymbol{\phi}_j = \langle \boldsymbol{\phi}_i, \boldsymbol{\phi}_j \rangle =  \delta_{ij}.
\end{equation}
This defines the natural inner product for the Laplacian operator \eqref{eq:VertexLaplacian}, which is the  operator governing the linearised dynamics of the transport in the maze.  The Laplacian \eqref{eq:VertexLaplacian} is positive semi-definite, with a single zero eigenvalue corresponding to a constant eigenvector representing steady state concentrations as $t \to \infty$. 
%Since the underlying maze has no symmetries, we find that all the %eigenvalues are distinct. 

%  we will use, and we see that the eigenvectors of the vertex Laplacian are orthogonal with respect to this inner product.
 % \begin{equation}
%     \hat{\Gamma}_t = \frac{1}{4} \bar{\Gamma} \hat{\Gamma}_{xx},
% \end{equation} 
% When we solve this equation on the graph which represents the maze, this second derivative will become our MFD Laplacian operator. 
 
%Now we will demonstrate that the eigenvectors of the (non-symmetric) vertex Laplacian are orthogonal with respect to a suitable inner product. 
 % For a vector $\boldsymbol{\Gamma}(t)$ defined on the vertices of the graph, the linearised problem is given by

\begin{figure}
    \centering
    \includegraphics[width=0.4\textwidth]{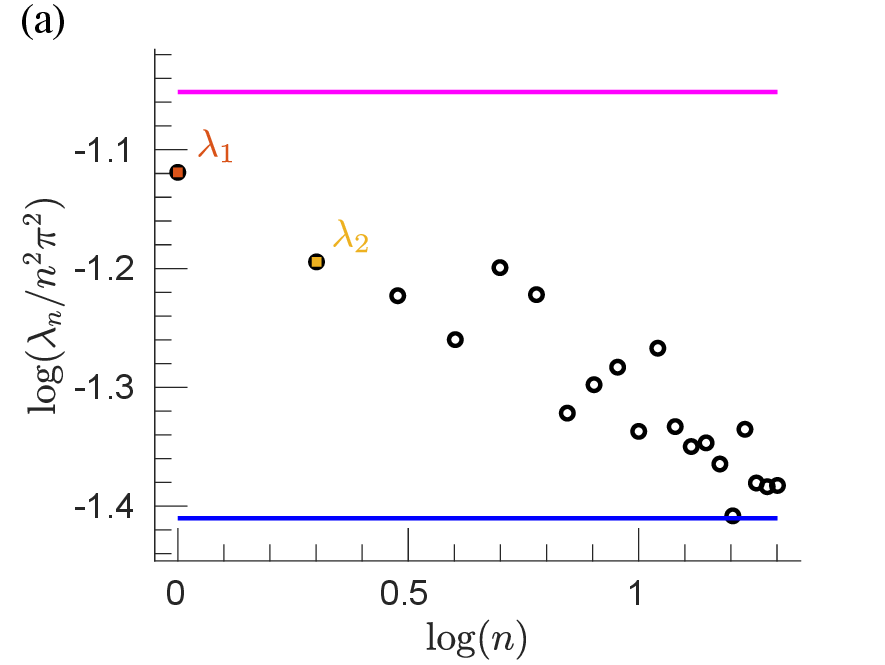}~~~~~~
    \includegraphics[width=0.4\textwidth]{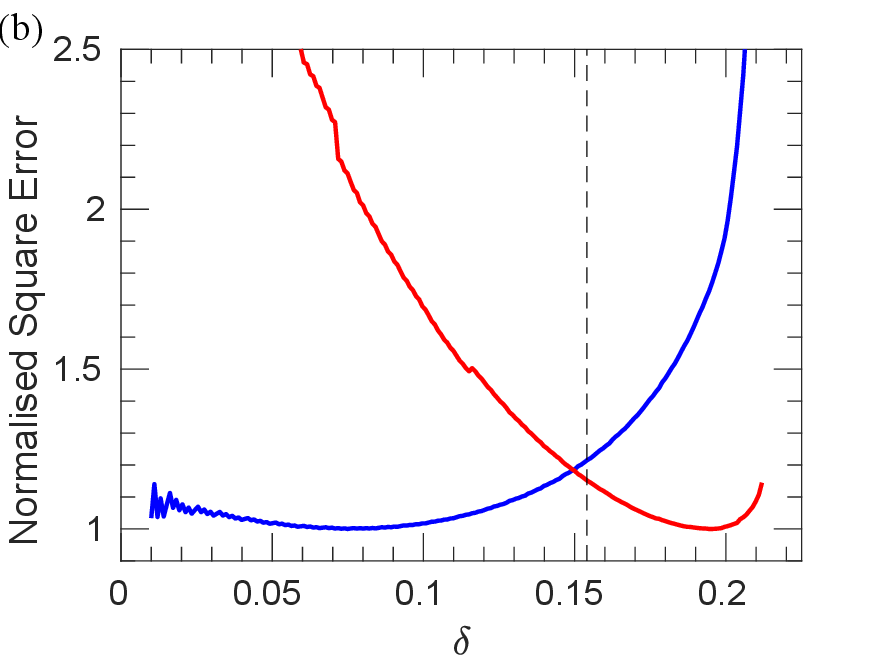}

    \caption{(a) The first 20 eigenvalues of the Laplacian (\ref{eq:VertexLaplacian}) (circles). In magenta is the approximation $\pi^2n^2/L_W$, where $L_W=3.36$ is the nondimensional length of the solution path. In blue is the approximation $\pi^2n^2/L_T$, where $L_T= 5.07$ is the sum of the lengths of all the branches in the maze. The first expression approximates the dominant eigenvalues well, but the higher order modes are much better approximated by the second expression.
    (b) Results of the parameter optimisation. The blue line shows $E_{sp}(\delta)/\min E_{sp}(\delta)$ calculated from \eqref{eq:ErrorSolutionPath}, the normalised square error between the front location in the experiment and simulation along the solution path. The red line shows $E_{pb}(\delta)/\min E_{pb}(\delta)$ calculated from \eqref{eq:Errorlateral}, the normalised square error between experiment and simulation for the  location of the fronts in the lateral branches. Both error calculations give  minima which are close to each other. The vertical dashed line represents the minimum of the sum of the two normalised data sets, which is the optimal value we use for $\delta$ at 0.15.}
    \label{fig:Modes}
\end{figure}

 We can use a truncated sum of the first $N$ eigenvectors to construct an approximate solution for the vector of surfactant concentrations defined on the vertices of the graph,
 \begin{equation}
     \boldsymbol{\Gamma}(t) = \sum_{i=0}^{N} A_i\boldsymbol{\phi}_i e^{-\lambda_i t},
 \end{equation} where the coefficients $A_i$ are the amplitudes of the eigenvectors $\boldsymbol{\phi}_i$, and $\lambda_i$ are the corresponding eigenvalues of \eqref{eq:VertexLaplacian}. Utilising the orthogonality of these eigenvectors with respect to this inner product we can project an initial concentration profile $\boldsymbol{\Gamma}(0)$ onto this basis. The amplitudes for the $i$th mode will be given by
 \begin{equation}
     A_i = \langle \boldsymbol{\phi}_i, \boldsymbol{\Gamma}(0) 
     \rangle.
 \end{equation}
 We find the modes look similar to cosine waves through the solution path of the maze. The gradients of the cosine waves have discontinuities at the locations of junctions, which is to be expected due to the internal boundary condition \eqref{eq:FluxCond}. 
The first 20 eigenvalues are plotted in figure \ref{fig:Modes}(a).

\subsection{Approximation of transport with the linear modes}\label{sec:Opt}

We can capture qualitatively the main behaviours observed in the experiment (maze solving, receding in dead-end sections) with only three eigenmodes. We are however unable to capture the initial conditions even approximately with only three modes. To replicate the key behaviours we therefore need to adjust $\delta$ and $L_0$. Using a fitting parameter $\Phi$, we replace $L_0$ with $\Phi L_0$. 
We perform a parameter sweep over $1 \leq \Phi \leq 40 $ and $0.01\leq \delta \leq 0.1$ to find a minimum error between experiment and simulation using the same error quantification as in section \ref{sec:ErrorComparison}. We find a minimum of the combined data sets at $\delta = 0.03$ and $\Phi = 25$. 
% The dominant three modes are plotted in figure \ref{fig:Modes}(a). The eigenvalues are plotted in figure \ref{fig:Modes}(b).

% \begin{figure}
%     \centering
%     \includegraphics[width=0.49\textwidth]{fig4a.eps}
%     \includegraphics[width=0.49\textwidth]{SupplementFig5EigvalsBest.eps}
%     \caption{A figure showing the three modes that optimise the scheme presented in section \ref{sec:Opt}. In (a) we present the components of three dominant modes along the solution path through the maze. These three modes alone can reproduce both the maze solving behaviour and receding in dead end sections. The thick vertical black lines represent the entrance and exit to the maze, with the components to the left and right representing IDD $0$ and $37$ respectively. The dashed vertical lines represent the locations of junctions within the maze. In (b) we plot the first 20 eigenvalues of the discrete Laplacian in circles. In magenta is the approximation $\pi^2n^2/L_M$, where $L_M$ is the length of the solution path. In blue is the approximation $\pi^2n^2/L_T$, where $L_T$ is the sum of the lengths of all the branches in the maze. The first expression approximates the dominant eigenvalues well, but the higher order modes are much better approximated by the second expression.}
%     \label{fig:Modes}
% \end{figure}

\section{Experimental data}
\label{sec:ExtractData}

Experimental data to compare with the model predictions were obtained through image analysis of the video of the experiment \cite{temprano2018soap}. The video of the maze experiment was converted into a time series of $N+1$ JPEG images, with each image taken at regular times $T_i$, where $i=0,1,2,\dots,N$. The time $T_0$ is the start time and the corresponding image is given in \fig~\ref{fig:numbscheme} (left), and $T_N$ is the completion time (chosen when the dye front just enters the outlet branch 37). We imported these images into MATLAB as matrices of dimension $h\times v \times 3$ where $h$ and $v$ are the number of rows and columns of pixels constituting the image, and for each pixel, the three numbers in the third index take integer values from $0$ to $256$, corresponding to the 8-bit light intensity value of the red, green and blue components at this pixel. We normalise these data by dividing by $256$, such that ideal white is created by the $3$-tuple $ [1,1,1]$, and ideal red by $[1,0,0]$. 

In each image of the time series, we average the pixel values in the transverse direction across the branches, for every branch in the maze, in order to obtain a 1D network of pixels representing the maze.
For instance, let us say a horizontal (east-west orientation) section of a branch in a given image has $P_w$ pixels in the east-west direction and $P_l$ pixels in length in the north-south direction, then its image is a $P_w \times P_l \times 3$ matrix $\mathsf{P}$. We transform this matrix into a $P_l \times 3$ matrix $\mathsf{p}$ defined by
\begin{equation}\label{eq:HSection}
   \mathsf{p}(j,k) =  \frac{1}{P_w}\sum_{i=1}^{P_w} \mathsf{P}(i,j,k).
\end{equation} 
Likewise, for a vertical section of a branch, its image is a $P_l \times P_w \times 3$ matrix, which we transform by
\begin{equation}\label{eq:VSection}
   \mathsf{p}(i,k) =  \frac{1}{P_w}\sum_{j=1}^{P_w} \mathsf{P}(i,j,k).
\end{equation} 
At corners we neglect the transverse averaging, and instead trace a line of pixels diagonally from the middle of one branch section incident to the corner, indexing this line with the first index of a matrix $\mathsf{c}(i,k)$, where $k$ are once again the $3$ RGB components. Junctions are taken to belong wholly to the branch nearest the inlet.
We then concatenate the matrices of the form \eqref{eq:HSection} and \eqref{eq:VSection}, and corner sections. For example, the $i$th branch of a maze at time $T_j$ consists of first a horizontal section, which we turn into a matrix of form \eqref{eq:HSection}, which we call $\mathsf{p}_{h1}$, followed by a corner matrix $\mathsf{c}_1$, followed by a vertical section of the form \eqref{eq:VSection}, which we call $\mathsf{p}_{v1}$, followed by a corner $\mathsf{c}_2$, and so on; the matrix representing the branch is then
\begin{equation}\label{eq:B}
    \mathsf{B}_i^{T_j} = \begin{bmatrix} \mathsf{p}_{h1} & \mathsf{c}_1 & \mathsf{p}_{v1} & \mathsf{c}_2 & \mathsf{p}_{h2} & \mathsf{c}_3 & \mathsf{p}_{v2} & \dots
    \end{bmatrix}^{\top}.
\end{equation} 
The image of all 36 branches in the maze are thus transformed into matrices of the form \eqref{eq:B}, where the first index represents the pixel location along the branch with its index increasing in the direction away from the junction nearest to the inlet, and the second index gives the transverse averaged pixel intensities at that location. This operation was repeated for each image in the time series from $T_0$ to $T_N$.  

For each image in the time series, we compute upper and lower bounds for the locations of the fronts of exogenous surfactant in each branch, as a measure of the spread of the front. We define the upper bound $\Lambda^u$ as the pixel with the smallest index which has both green and blue components above $180$, and the lower bound $\Lambda^l$ as the pixel with the minimum index where both green and blue were above $120$. The difference in these values represent approximately a quartile of colour intensity. For larger values than 180, or smaller values than 120, we find a smooth approximation of the front location is unobtainable due to noise. Indeed, in both the video images, and transverse averaged 1D network, the pixel components creating the white of the milk and the red of the dye are noisy, such that any given pixel usually has RGB components in the range $ [0.9,0.9,0.9] \leq I_{milk} \leq [1,1,  1]$ and $ [0.9,0,0] \leq I_{red} \leq [1,0.1,  0.1]$. If the number of rows in matrix $\mathsf{B}_i$ is $P_B$, then these locations are
\begin{equation}\label{eq:UpperBound}
    \Lambda^{u}_i(T_j) = \max_{D:(B_i^{T_j}(D,2)>180)\cup(B_i^{T_j}(D,3)>180)}   \frac{\AR}{P_B} L_i, \ \text{and} \ \Lambda^{l}_i(T_j) = \max_{D:(B_i^{T_j}(D,2)>120)\cup(B_i^{T_j}(D,3)>120)} \frac{\AR}{P_B} L_i.
\end{equation}
% and
% \begin{equation}\label{eq:LowerBound}
%     \Lambda^{l}_i(T_j) = \max_{D:(B_i^{T_j}(D,2)>120)\cup(B_i^{T_j}(D,3)>120)} \frac{\AR}{P_B} L_i.
% \end{equation}
The difference  $|\Lambda^{l}_i(T_j)-\Lambda^{u}_i(T_j)|$ gives us an experimental error for the exogenous front location. We define $\Lambda_i(T_j)$ as the mean of the upper and lower bounds
\begin{equation}\label{eq:ExpFrontLocation}
    \Lambda_i(T_j) = \frac{1}{2} \left(\Lambda^{u}_i(T_j) +\Lambda^{l}_i(T_j)  \right).
\end{equation} The initial location of the unique exogenous front as used in \eqref{eq:initcond3}, \eqref{eq:FrontInitCond}, and \eqref{eq:L_0} is given by $\Lambda_1(0)=x_f=0.032$.

\section{Comparison of the model predictions with the experiment}
\label{sec:ErrorComparison}

We compare the predictions for the time evolution of the fronts of the exogenous surfactant $\widetilde{\Lambda}_i$, obtained from solving the MFD formulation \eqref{eq:NumericalScheme} and \eqref{eq:NumericalFrontLocation}, with the experimental data $\Lambda_i(T_j)$. We analyze two distinct behaviours. The first is the front location of exogenous surfactant along the solution path, and the second is the late time behaviour of the front locations in the lateral branches. Hence, we define two separate error quantification procedures for the two behaviours and perform a parameter sweep over $\delta$ to find the values which minimise both errors.  

\subsection{Minimisation of the error along the solution path}
\label{subsec:ErrorSolutionPath}

We define $E_{sp}(\delta)$ as the difference between simulation and experiment for the front location along the solution path,  defined by branches $1$, $3$, $5$, $7$, $9$, $12$, $14$, $18$, $22$, $30$, $32$, $34$ and $36$ (\fig~\ref{fig:numbscheme}, left).
We choose a set of $M$ equally spaced points along the solution path, and for each point $\chi_i$ find the time taken $\widetilde{\tau}_i$ for the exogenous surfactant front to arrive in a simulation run with a given $\delta$. Then, we use the data set obtained from the experiment of exogenous front locations to find an analogous data set from the experiment. The data set found in \ref{sec:ExtractData} maps  points equally spaced in time $T_j$ to the spatial location of the exogenous front in branch $i$, $\Lambda_i(T_j)$. Using MATLAB\textsuperscript{\textregistered}'s `interp1' function we interpolate between these data points to find the data set mapping the same set of spatial points $\chi_i$ along the solution path to the time of arrival at those points $\tau_i$.

We  obtain two normalised, comparable data sets from simulation and experiment $\widetilde{\tau}_i/\widetilde{\tau}_M$ and $\tau_i/\tau_M$, respectively, for $i=1,2 ,\dots M$. We impose  that experiment and simulation start at the same time, but we allow the finish time to vary by scaling the simulation data set by a  parameter $0.5\leq\tau_{end}\leq 1.5$. This secondary parameter  provides a slightly better fit, but has negligible impact on the fundamental dynamics underpinning the model. To obtain a positive definite error 
% quantification, we subtract simulation data  from the experimental data, square the result, and then sum these values together. We perform a parameter sweep over $0.5 \leq \tau_{end} \leq 1.5$ to find the minimum error $E_{sp}(\delta)$
we compute
\begin{equation}\label{eq:ErrorSolutionPath}
    E_{sp}(\delta) = \min_{0.5\leq \tau_{end} \leq 1.5 }\sum_{i=1}^M \left(\frac{\tau_i}{\tau_M} - \tau_{end}\frac{\widetilde{\tau}_i(\delta)}{\widetilde{\tau}_M(\delta)} \right)^2.
\end{equation}
We compute  $E_{sp}(\delta)$ for a range of values $0< \delta< 1$. The normalised error $E_{sp}(\delta)$ is shown  in \fig~\ref{fig:Modes}(b) (blue). The minimum of  $E_{sp}(\delta)$ is found at  $\delta \approx 0.075$.

\subsection{Minimisation of the error in the lateral branches}

We  quantify the error $E_{pb}(\delta)$ between experiment and simulation for the front location in the lateral branches. We consider branches, $I = \{2,4,6,8,10,11,13,15,19,23,24,31,33\}$ (\fig~\ref{fig:numbscheme}, left). We neglect branch $35$ where the experimental front does not penetrate. 
We record the location of the front inside each of the lateral branches in the set $I$ at the set of time points $1\leq j \leq M$ described in subsection \ref{subsec:ErrorSolutionPath}, in both simulation $\widetilde{\Lambda}_i(\widetilde{\tau}_j)$ and experiment ${\Lambda}_i({\tau}_j)$. We  compute the error as
\begin{equation}\label{eq:Errorlateral}
    E_{pb}(\delta) = \sum_{i \in I} \sum_{j=1}^M \frac{ \left( \Lambda_i(\tau_j) -\widetilde{\Lambda}_i(\widetilde{\tau}_j)\right)^2 }{L_i^2}.
\end{equation}
The normalised error $E_{pb}/\min E_{pb}$ is plotted  in \fig~\ref{fig:Modes}(b) (red) for a range of $\delta$. The  minimum is at $\delta \approx 0.195$.

\subsection{Optimal value for the endogenous concentration}

The errors $E_{sp}(\delta)$ and $E_{pb}(\delta)$, defined in \eqref{eq:ErrorSolutionPath} and \eqref{eq:Errorlateral} respectively, give well-defined minima with different optimal values for $\delta$. We added these normalised data sets together, and found a combined minimum  at $\delta \approx 0.15$ with $\tau_{end}\approx 1.24$, which are the values used in the simulation results shown in this study. We note that all values $0.02\leq \delta \leq 0.175$ would give a reasonable fit between the model predictions and the experiments along the solution path, and all values $0.125 \leq \delta \leq 0.21$ would give a reasonable fit of the dynamics in the lateral branches. 
% This shows that the model is not very sensitive to the fitting parameter $\delta$. %he chosen value for our main fitting parameter $\delta \approx 0.15$ is  a physically sensible experimental value, as would be expected for the level of endogenous surfactant compared to exogenous surfactant where the endogenous surfactant is likely to come from environmental contamination.

\section{Dynamics in lateral branches}

\begin{figure}[h]
		\centering
			\includegraphics[width=0.32\textwidth]{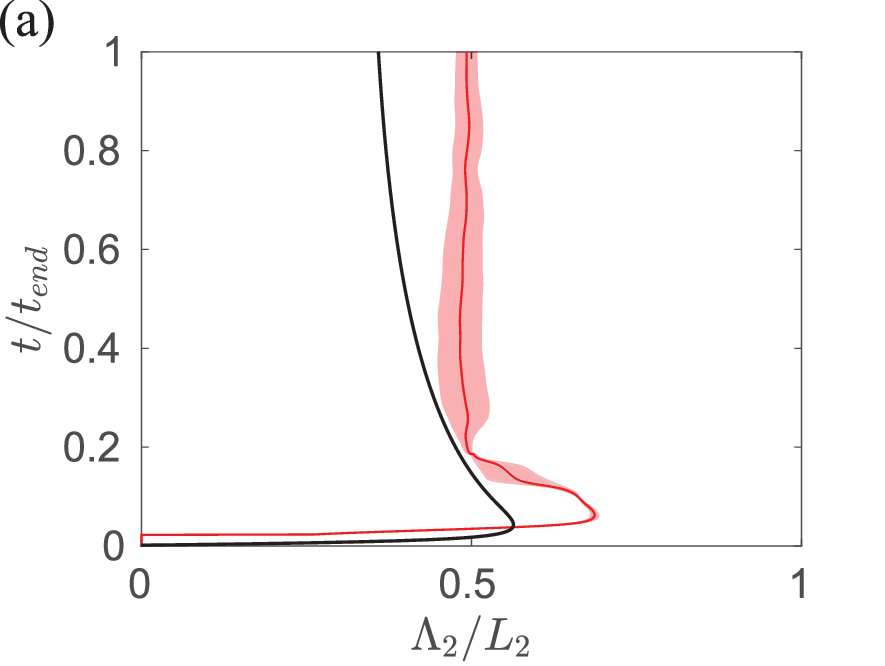}
			\includegraphics[width=0.32\textwidth]{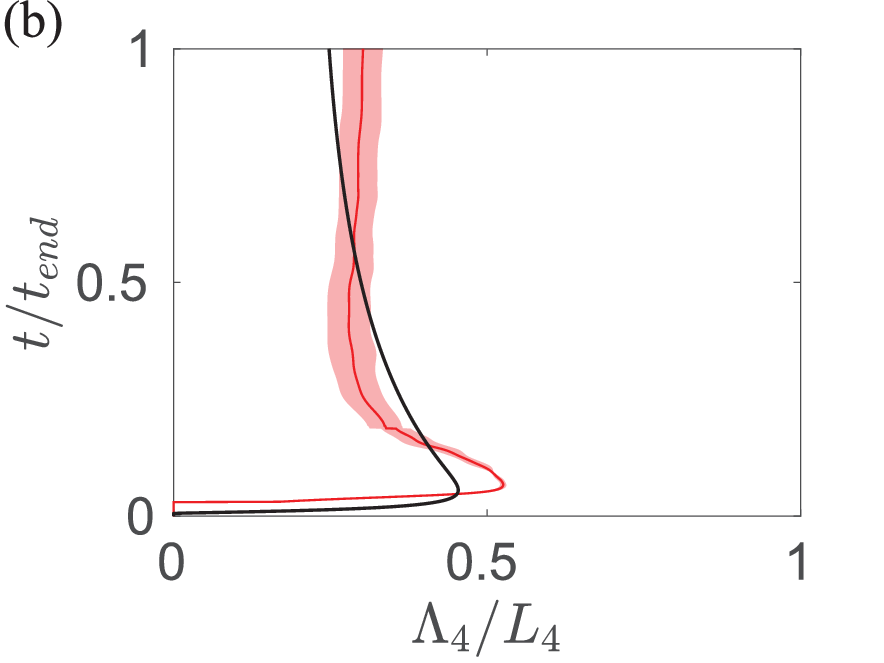}
			\includegraphics[width=0.32\textwidth]{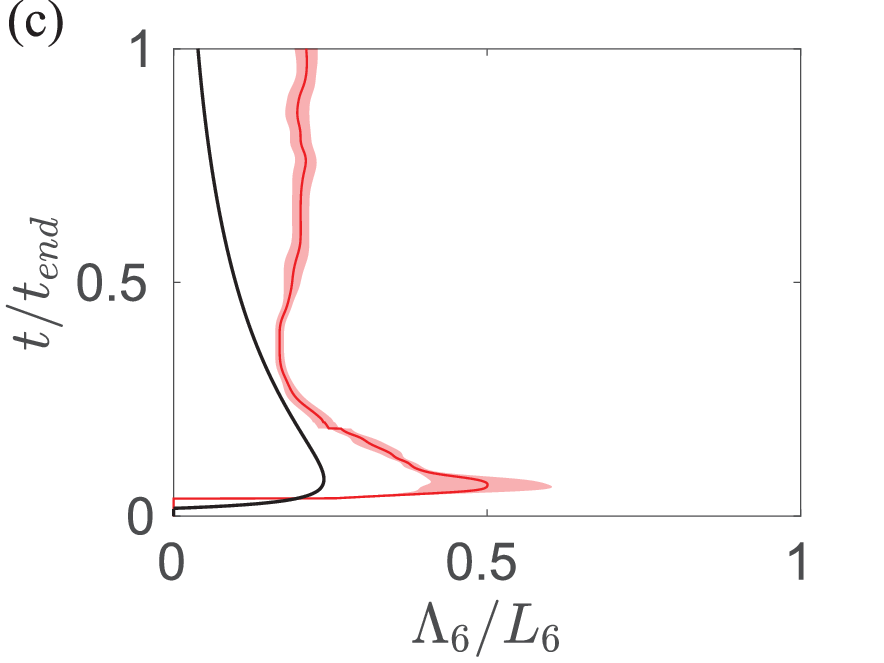}
			\includegraphics[width=0.32\textwidth]{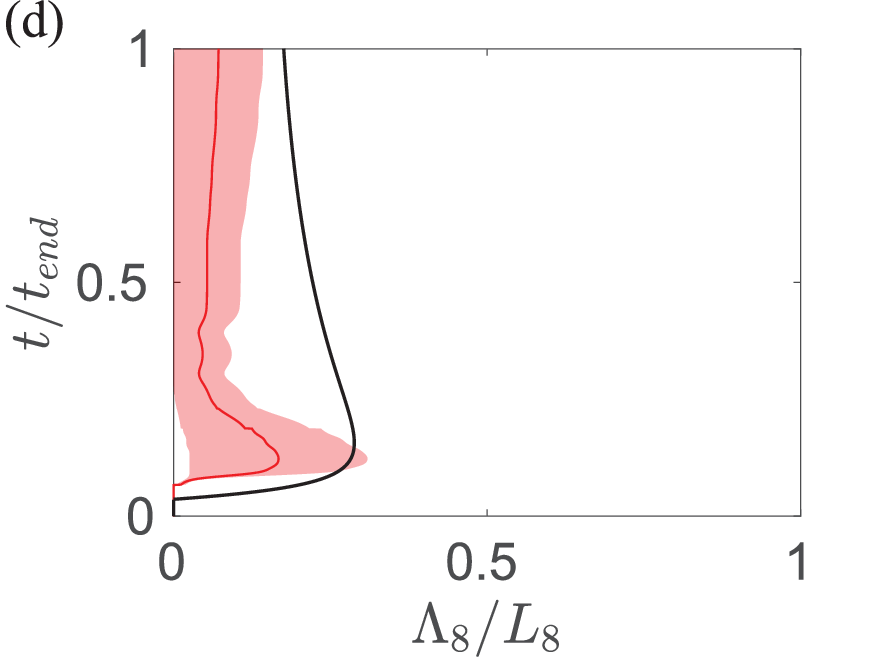}
		\includegraphics[width=0.32\textwidth]{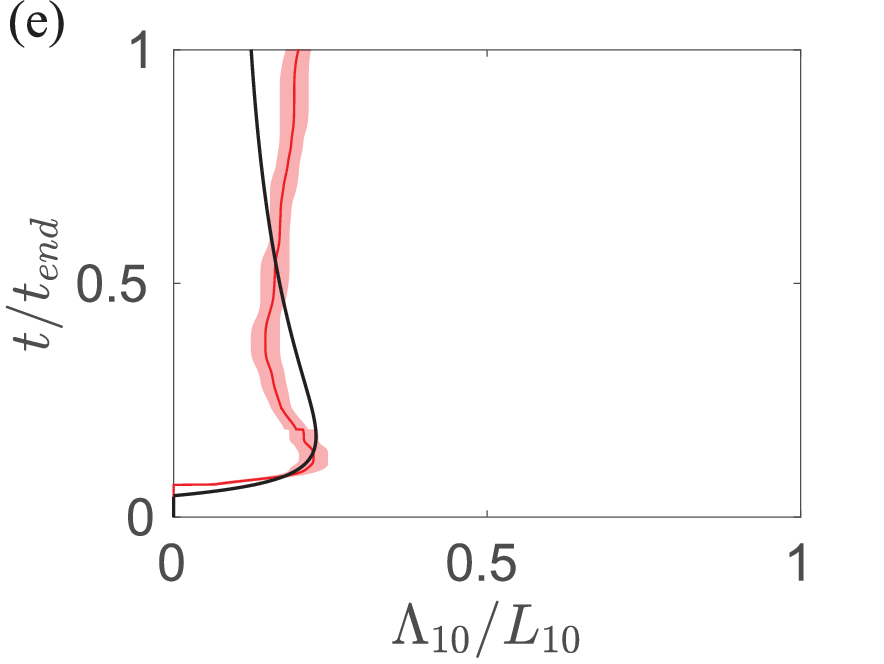}
			\includegraphics[width=0.32\textwidth]{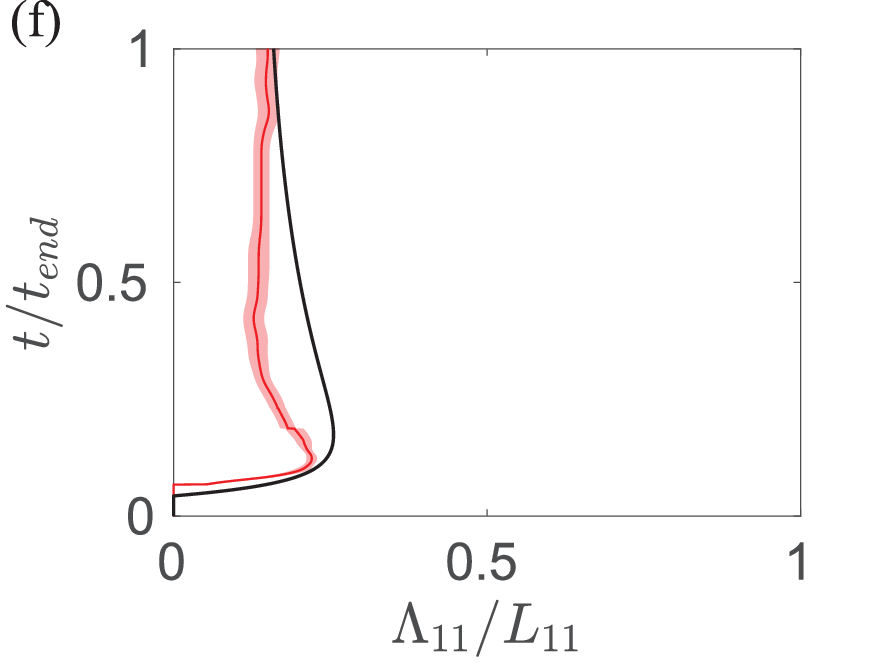}
			\includegraphics[width=0.32\textwidth]{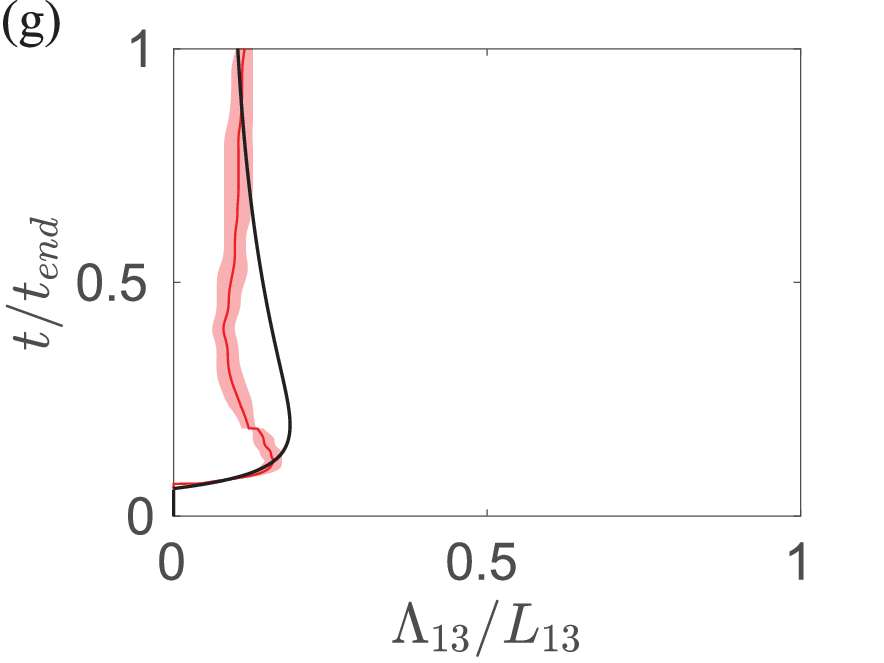}
			\includegraphics[width=0.32\textwidth]{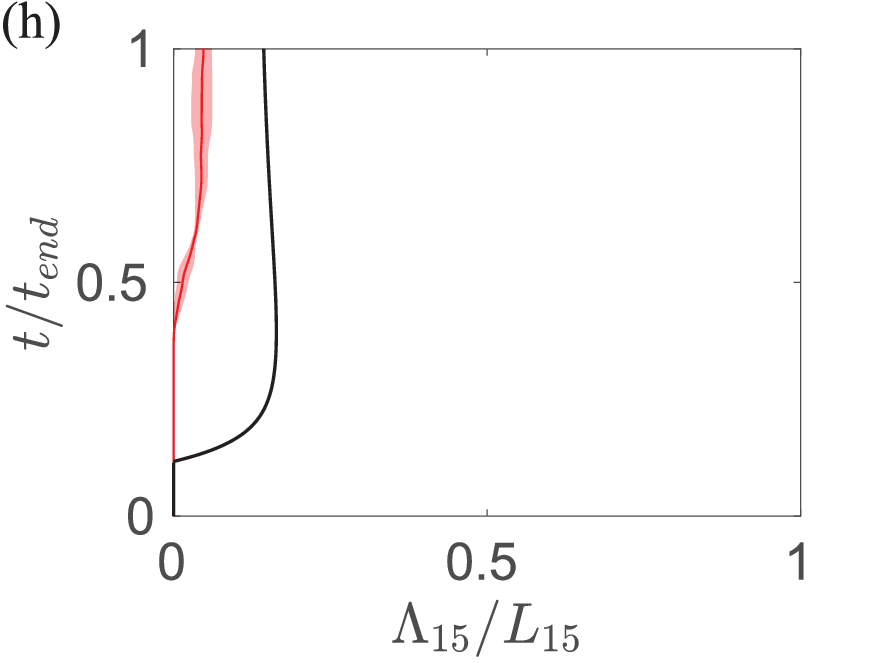}
		\includegraphics[width=0.32\textwidth]{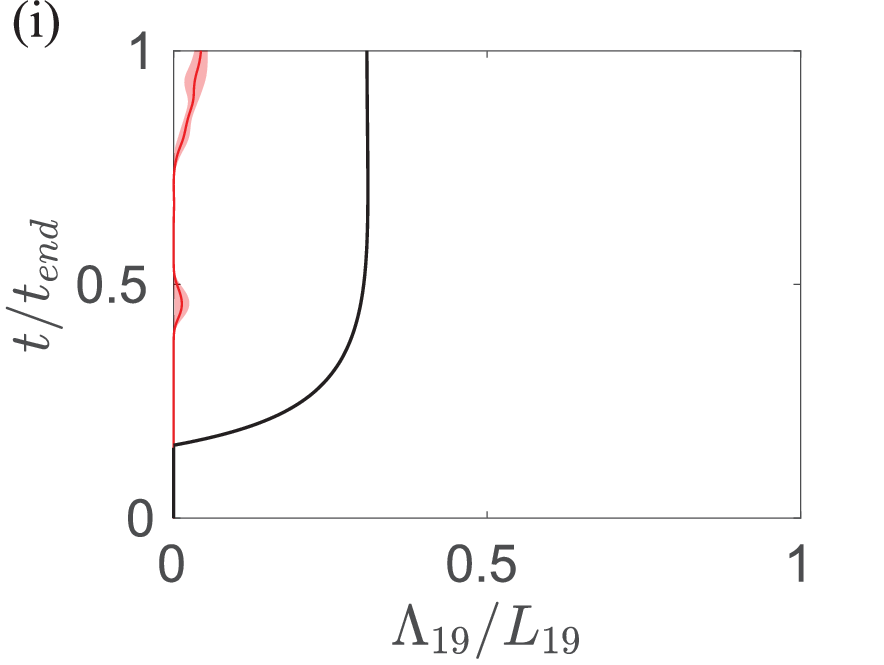}
		\caption{
 		Evolution of the front in time in the first $9$ lateral branches for both the optimal prediction from the nonlinear model (black curve,  $\delta=0.15$) and the experimental data (red curve) \eqref{eq:ExpFrontLocation}. 
 		The red shaded area shows the experimental error \eqref{eq:UpperBound} for the front location. 
  		We observe receding behaviour in both simulation and experiment as the non-monotonic behaviour of surfactant concentration in the early branches of the maze cause a Marangoni flow in the opposite direction in these branches. The receding in the experiment seems to occur over shorter timescales than we achieve in the simulation.
}
		\label{fig:SideBranches}
	\end{figure}

In addition to the results  presented in \fig~3(a) (inset) in the main paper, we show the front dynamics in the first $9$ lateral branches in the experiment and in the simulation using the optimal value of $\delta$ calculated in section \ref{sec:ErrorComparison}, \fig~\ref{fig:SideBranches}. We can observe the same receding effect of the front locations in the simulation and in the experiment. This is explained physically by the finite mass of surfactant available from the inlet branch (in contrast to a fixed concentration at the inlet) spreading through the solution path causing non-monotonic behaviour of the surfactant concentration closer to the inlet. The concentration increases substantially initially in the first lateral branches of the solution path, but then decreases at later times. This late time behaviour causes a receding Marangoni flow in the opposite direction in branches $2$, $4$, $6$, $8$, $10$, $11$, $13$, and $15$.

% \begin{figure}
%     \centering
%     \includegraphics[width=0.4\textwidth]{OptimisationFig2.eps}
%     \caption{Graph showing the results of the parameter fitting scheme. The blue line shows $E_{sp}(\delta)/\min E_{sp}(\delta)$ calculated from \eqref{eq:ErrorSolutionPath}, the normalised square error between the front location in the experiment and simulation along the solution path. The red line shows $E_{pb}(\delta)/\min E_{pb}(\delta)$ calculated from \eqref{eq:Errorlateral}, the normalised square error between experiment and simulation for the  location of the fronts in the lateral branches. Both error calculations give  minima which are close to each other. The vertical dashed line represents the minimum of the sum of the two normalised data sets, which is the optimal value we use for $\delta$ at 0.154.}
%     \label{fig:OptFig}
% \end{figure}

\section{Two-dimensional dynamics}

We use COMSOL to simulate the spreading of surfactant in some of the individual branches, and use the particle-tracing feature to track the dynamics of the front of exogenous surfactant. We do this for the  section which consists of part of branches $7$ and $14$, and branches $8$, $9$, $10$, $11$, $12$ and $13$ (Fig.~\ref{fig:numbscheme}, left).
We solve the 2D equation
\begin{equation}
 \frac{\partial \Gamma}{\partial t} = f(\AR) \boldsymbol{\nabla} \cdot (\Gamma \boldsymbol{\nabla}\Gamma)
\end{equation} 
in this section, implementing no flux boundary conditions on all boundaries except at the inlet ($\Gamma=1$) and outlet $\Gamma=\delta$.
The initial conditions  are $\Gamma =  (1-\delta)\cos{(\pi x/2)}/2 +(1+\delta)/2$ for $0\leq x_7\leq2$, and $\Gamma = \delta$ everywhere else.

% \begin{figure}[]
% 		\centering
% 		\begin{subfigure}{0.33\textwidth}
% 			\includegraphics[width=\textwidth]{FirstBranch.jpg}
% 			\caption{}
% 		\end{subfigure}
% 		\begin{subfigure}{0.48\textwidth}
% 			\includegraphics[width=\textwidth]{MazeBranchdelta0p29R0p25.png}
% 			\caption{}
% 		\end{subfigure}
% 		\begin{subfigure}{0.46\textwidth}
% 			\includegraphics[width=\textwidth]{MidBranchesExp.jpg}
% 			\caption{}
% 		\end{subfigure}
% 		\begin{subfigure}{0.48\textwidth}
% 			\includegraphics[width=\textwidth]{MidBranchEnddelta0p7.png}
% 			\caption{}
% 		\end{subfigure}
% 		\caption{}
% 		\label{fig:2DDynamics}
% 	\end{figure}

% We simulate many complex two dimensional features observed in the maze experiment like corner-hugging, as presented in Fig.~4 in the main paper. This corner-hugging behaviour is due to the way that level curves of $\Gamma$ travel along long thin branches. They travel as parallel lines, and so when they travel round a sharp corner, the level curves becomes more dense around the inside of the corner increasing the gradient of $\Gamma$, and they spread our around the outside of the corner, decreasing this gradient.

\bibliographystyle{apsrev4-2}
\bibliography{Refs}